\documentclass{article}

\usepackage[english]{babel}

\usepackage[a4paper]{geometry}
\usepackage{bm}
\usepackage{colortbl}

\usepackage{caption}
\usepackage{latexsym, amssymb, amscd, amsxtra}
\usepackage{graphics, graphicx, color, subcaption, float, multirow}
\usepackage[dvipsnames]{xcolor}
\usepackage{epsfig}
\usepackage[colorlinks=true, allcolors=blue]{hyperref}
\usepackage[open, openlevel=3, atend, numbered]{bookmark}

\usepackage{amsthm}
\theoremstyle{definition}
\newtheorem{definition}{Definition}
\newtheorem{prop}{\protect\propositionname}
\newtheorem{rem}{\protect\remarkname}

\providecommand{\lemmaname}{Lemma}
\providecommand{\propositionname}{Proposition}
\providecommand{\remarkname}{Remark}
\providecommand{\theoremname}{Theorem}

\usepackage{titlesec}
\titleformat{\paragraph}
{\normalfont\normalsize\bfseries}{\theparagraph}{1em}{}
\titlespacing*{\paragraph}
{0pt}{3.25ex plus 1ex minus .2ex}{1.5ex plus .2ex}

\usepackage[shortlabels]{enumitem}
\usepackage{authblk}

\usepackage{natbib}
\newcommand{\petit}{\hspace{-0.1cm}}

\newcommand{\noprint}[1]{}

\title{Curvature and Torsion estimation of 3D functional data: A geometric approach to build the mean shape under the Frenet Serret framework}
\author[1,2]{Juhyun Park} 
\author[1,2,3]{Nicolas Brunel}
\author[1]{Perrine Chassat}

\affil[1]{\small LaMME, Universit\'e Paris-Saclay, CNRS, France} \affil[2]{\small ENSIIE, \'{E}vry, France} \affil[3]{\small Quantmetry, Paris, France}

\begin{document}
\maketitle

\begin{abstract}
The analysis of curves has been routinely dealt with using tools from functional data analysis. However its extension to multi-dimensional curves poses a new challenge due to its inherent geometric features that are difficult to capture with the classical approaches that rely on linear approximations. We develop an alternative characterization of a mean that reflects shape variation of the curves. Based on a geometric representation of the curves through the Frenet-Serret ordinary differential equations, we introduce a new definition of mean curvature and mean torsion, as well as mean shape through the notion of mean vector field. This new formulation of the mean for multi-dimensional curves allows us to integrate the parameters for the shape features into the unified functional data modelling framework. We formulate the estimation problem of the functional parameters in a penalized regression and develop an efficient algorithm. We demonstrate our approach with both simulated data and real data examples. 
\end{abstract}

\section{Introduction}

%The rise of data in the form of curves in multivariate fashion in recent years has seen the surge of interest in the multivariate functional data analysis. For example, the registration problem of multivariate growth curves \citep{Carroll2020}, multivariate functional mixed modelling for spectroscopy data \citep{Zhu2017} and the cross-correlation for multivariate human gesture tacking problem \citep{Zhang2020} demonstrate the diversity of the domains of interest. 

\begin{figure}[h]
    \centering
    %\begin{subfigure}[b]{1\textwidth}
    %    \centering
    \begin{tabular}{lr}
        \hspace*{-1cm}\includegraphics[width=0.42\textwidth]{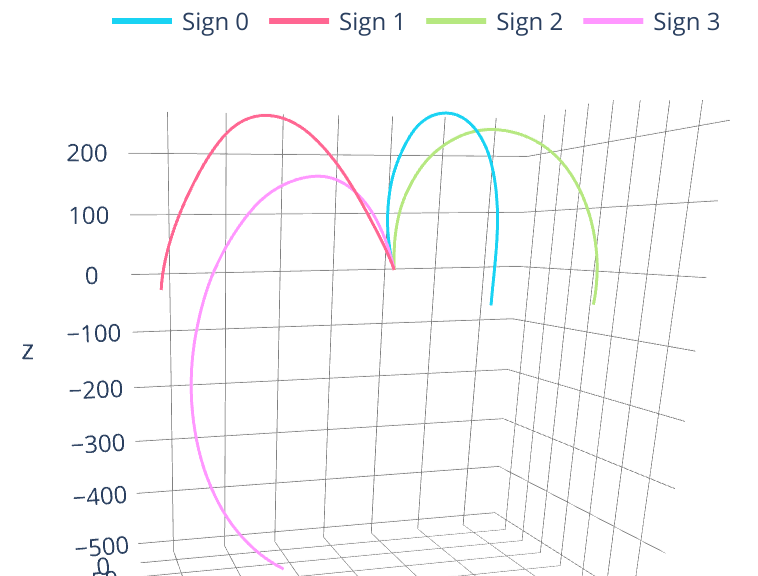} &
        \includegraphics[width=0.5\textwidth]{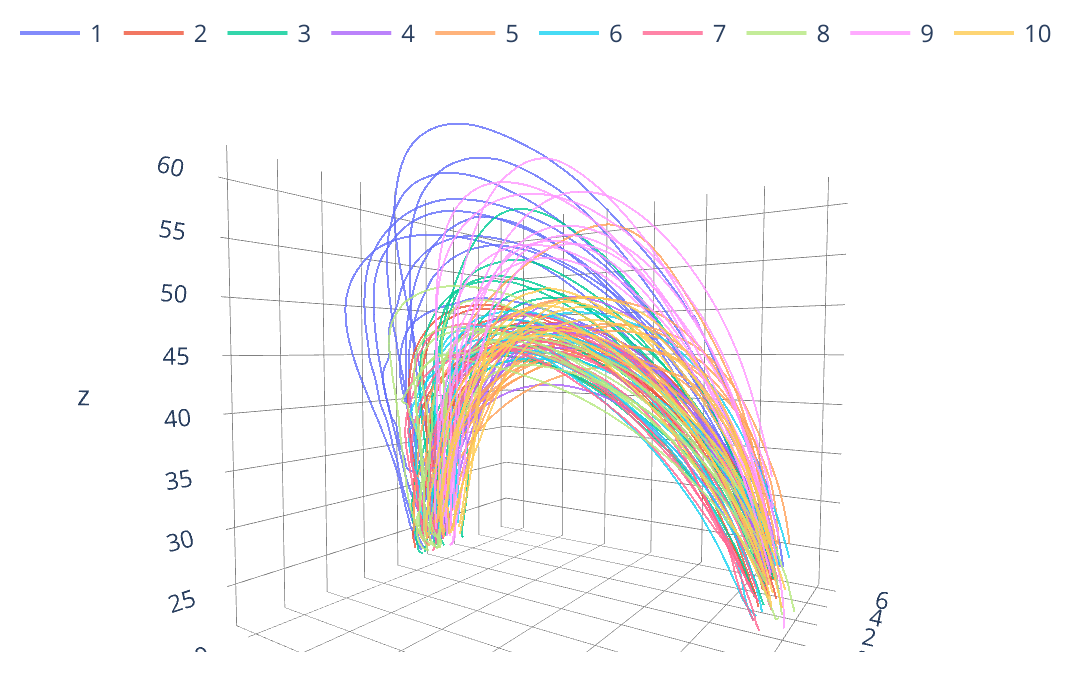}
    \end{tabular}
    \caption{Motion signals: sign "FLY" (left) and hand movements from biomechanical experiments in \citet{RaketMarkussen2016}. (right). On the right, different colours represent different subjects.}
    \label{fig:realdata}
\end{figure}

We consider the problem of analyzing a set of three-dimensional curves in $\mathbb{R}^3$ in the spirit of functional data analysis. A typical example would be the recordings of spatial coordinates for tracking movements of body parts or objects (e.g., \citet{FlashHogan1985}). 
Our motivating example deals with the movements signals automatically captured by a motion capture system by the company MOCAPLAB\footnote{https://www.mocaplab.com/fr/}. Among the many fields of exploration of motion capture is the very specific field of sign language involving movements of the body, hands, fingers, face and eyes and achieve a capacity for expression as rich and structured as that offered by speech \citep{MocapLSF}. An example of the sign signals as well as biomechanical experimental data \citep{RaketMarkussen2016} is depicted in Figure~\ref{fig:realdata}. These types of movement are challenging to study as they are meaningful but are difficult to characterize without specific knowledge in the field. Our aim is to develop statistical tools to extract ``primitives" or a template specific to the nature of the signals studied. 
%These may correspond to the notion of a template or a mean, which we will make it clear later.

It is of scientific interest to analyze three-dimensional curves in terms of curvature and torsion \citep{Lewiner2005,Sangalli2009, Kim2013}. Indeed, the geometry of the trajectories of movement have physical significance: curvature and torsion characterize this geometry and can provide insightful summaries of kinetic curves to scientists. 
%As these parameters have a physical meaning, such analysis could offer more interpretable summaries. 
This is a challenging task as curvature and torsion depend on higher order derivatives and their estimation from real data (even with a low noise) can be very unstable. Hence, the focus has been more on estimating derivatives in nonparametric regression and the link to functional data is somewhat lost. 

On the other hand, certain geometric variation of curves, often in two or three-dimensions, is studied under shape analysis. A notion of shape is understood as what is left invariant under the actions of the rigid transformations of the Euclidean space, i.e., rescaling, translating and rotating. %The definition alone however is too general to be useful in statistical modelling considered in FDA. 
Viewing shapes as points on a manifold, shapes are formally defined as equivalence classes under some appropriate group actions. 
Several ways of constructing shape spaces (or feature spaces) have been proposed: discretization with landmarks based after discretization \citep{Dryden1998}, or infinite dimensional shape spaces \citep{Younes2010, Srivastava2011}. The classical statistical methodologies need to be adapted in order to deal with the non-Euclidean properties of shape spaces.
Such shape analysis requires a definition of distance (inducing a Riemannian structure for instance) between points on a manifold and the natural extension of the usual mean is defined as a Fr\'echet mean. 
Some of these ideas, such as the elastic shape analysis has been suggested for analyzing the variations of functional data, typically for the registration problem in one-dimensional curves \citep{kurtek2012statistical}. Its extension to multi-dimensional curves is found in \citet{SrivastavaKlassen2016}. Under this framework, the main task is to estimate a geodesic on the manifold to quantify similarity of shapes. Hence, the variation of shapes tends to be limited to deformations on a geodesic and a statistical problem is somewhat hidden in the optimization problems. Also, the link to physical parameters is lost.  

Our aim is to integrate both types of analysis in a unified framework to characterize a mean that respects the geometry of the curves and, at the same time, maintain the link to the physical parameters. 
We begin by treating multidimensional curves as a special instance of multivariate functional data. %\citep{Zhu2017}.
A standard assumption with functional data analysis (FDA) \citep{RamsaySilverman2005, FerratyVieu2006, WangMuller2016} is that there exists a common structure, often through a common mean and variance function, which then allows for a parsimonious decomposition of variability through functional principal component analysis.  
%Then variability decomposition around the mean through functional principal component analysis provides a parsimonious decomposition of the variability. 
This type of linear approximations is powerful as it allows us to naturally extend tools for univariate methods to multivariate ones \citep{ChiouChenYang2014, Happ2018}. These ideas were successfully applied to movements data \citep{Goldsmith2016, Backenroth2018}. 
Nevertheless, such analytic extension can also hide some important features in these types of multivariate functional data \citep{DaiGenton2018}.
% as multidimensional curves or trajectories \citep{DaiGenton2018}. 
By borrowing ideas from statistical shape analysis \citep{Dryden1998, Kim2021}, we develop an alternative characterization of the common structure that is linked to the common geometry of the curves, which can be viewed as a geometric mean.  
The notion of a geometric mean has been used in functional data analysis somewhat intuitively in defining {\it structural mean} in the presence of phase variation \citep{KneipGasser1992}, more generally with manifold structure \citep{ChenMuller2012}, and in detecting shape outliers in multivariate functional data \citep{DaiGenton2018}.
Recent developments in functional data analysis focus on generalizations beyond the Euclidean data by allowing for non-standard features such as data on a manifold or in a general metric space \citep{Lin2019, Dubey2019, Petersen2019b}. 
Our approach is complimentary, as our generalization is to facilitate the Euclidean data analysis by incorporating non-Euclidean features, towards enhancing its interpretability.
%On the other hand, we extend the Euclidean data analysis by incorporating non-Euclidean features, enhancing its interpretability.
%In our generalization, we introduce the notion of mean curvature and mean torsion as the mean parameter, which then defines a mean shape. This can be interpreted as a geometric mean of the curves. 
%We use some of non-Euclidean properties of a derived Riemannian structure in our development, but without the need of a Fr\'echet mean. %
%We use the geometric representation of curves based on the local orthonormal basis representation known as the Frenet frames. Under this framework, the estimation of the mean shape can be cast into the problem of an ODE estimation in Lie group. To accompany the new definition of the mean, we propose a statistical framework for inference and develop an efficient algorithm. 

%A natural strategy in shape analysis and non-Euclidean data analysis would be to define a proper space equipped with a good metric where one can define a Fr\'echet mean. We derive some of non-Euclidean properties in our formulation, but without the need of a Fr\'echet mean. 

We consider a new distance between curves that does not depend on the usual Cartesian coordinate system but uses a parameterization of the space of smooth curves based on a geometric curve representation. This representation provides a local orthonormal basis system and is shown to be related to the Frenet Ordinary Differential Equation (ODE). We treat this representation as a prototype of our statistical model and explicitly define shape variation and phase variation models under our new framework.
We show that the solution of the ODE, the Frenet paths, can be interpreted as a representative of the equivalence class or {\it shape}. 
Within this framework, we introduce a new definition of mean shape through the mean ODE (or flow). In particular, we introduce the notion of mean curvature and mean torsion within this framework and show that the estimation of parameters can be cast into the problem of an ODE estimation in a Lie group \citep{Hairer2006}. To accompany the new definition of the mean, we propose a statistical framework for inference and develop an efficient algorithm.  In general, the ODE estimation is a difficult problem (e.g., \cite{Ramsay2007}), especially when involving nonparametric estimation of time-varying parameters (e.g. \cite{Mueller2010, WuDing2014}), even  without the orthogonality constraint required in our formulation. As a by-product, our formulation offers a new solution to a non-trivial ODE inference problem. We refer readers to \citet{RamsayHooker2017} for recent development on data analysis with ODE models. 

The paper is organized as follows. Section~\ref{sec:preliminaries} introduces fundamentals of the curve representation and reviews related concepts from elastic shape analysis. Section~\ref{sec:FSmean} develops new characterizations of a mean shape under our statistical framework.  Section~\ref{sec:estimation} presents our estimation algorithms, followed by numerical studies in Section~\ref{sec:numeric}. Main proofs, additional derivations and background materials are given in the supplementary. Our code is made available in the Python package \texttt{FrenetSerretMeanShape} \href{https://github.com/perrinechassat/FrenetSerretMeanShape}{(https://github.com/perrinechassat/FrenetSerretMeanShape)}. 

\section{A geometric representation of curves} \label{sec:preliminaries}

We are interested in analysing a set of curves in $\mathbb{R}^3$ defined as functions $\{x: [0, T] \rightarrow  \mathbb{R}^3\}$. 
%Let us consider $N$ curves in $\mathbb{R}^{3}$ defined as functions $\mathcal{S}=\left\{ x_{1},\dots,x_{N}\right\}$ from $\left[0,T\right]$ to $\mathbb{R}^{3}$. 
In order to simplify the exposition, we assume that the curves are \emph{regular, i.e.,} of class $C^{r}, r\geq 3$ (w.r.t time $t$) and the time derivative $\dot{x}(t)$ never vanishes on $[0,T]$. For notation, we write $\|\cdot\|$ for Euclidean norm in $\mathbb{R}^p$ and $\|\cdot\|_2$ for $L_2$ functional norm.

\subsection{Shape of the curve} \label{sec:shape}

For a curve $x$, the arclength is defined as $s(t)=\int_{0}^{t}\Vert \dot{x}(u)\Vert _{2}\,du, t \in [0, T]$
and $s(T)=L$ is the total length of the curve $X=\left\{ x(t),\,t\in\left[0,T\right]\right\}$.
%(we assume that the curves $\mathbf{X}_{i}$ have the same length $L$ equals to 1, possibly after renormalization). 
The \emph{shape} of the curve $X: \:\left[0,L\right]\longrightarrow\mathbb{R}^{3}$ is the image of the function $x$, which satisfies $x(t)=X(s(t))$.
The derivation with respect to arclength $s$ is denoted with prime
i.e $Y'(s)=\frac{d}{ds}Y(s)$, whereas time differentiation
is always denoted by a dot. 
For convenience, we use the arclength parametrization but the shape function is preserved under different parametrization of the curve, as demonstrated in Section~\ref{sec:timewarping}.
Our interest is in characterizing the variation of the shape function $X$ in the population of curves $x$.

\subsection{Curvature and Torsion} \label{sec:curvature}

The curvature and torsion are geometric invariants of the curve, independent of the parametrization of a curve $X$. Moreover, they completely describe the local behaviour of the curve, in the sense that two curves with the same curvature and torsion are identical up to translation and rotation. 
That is, they are invariant under the action of rigid (Euclidean) motions. 
These functional parameters can be directly defined with extrinsic formulas as 
\begin{equation}
\kappa(s(t))=  \frac{\Vert \dot{x}(t)\times\ddot{x}(t)\Vert}{\Vert \dot{x}(t)\Vert^{3}}\,,\qquad
\tau(s(t)) = \frac{\langle \dot{x}(t) \times \Ddot{x}(t), \dddot{x}(t) \rangle}{|| \dot{x}(t) \times \Ddot{x}(t) ||^2} \,.
% \tau(s(t))=  \frac{\langle \dot{x}(t)\times\ddot{x}(t),\dddot{x}(t)\rangle }{\Vert \dot{x}(t)\Vert _{2}^{3}} \,.
\label{eq:ExtrinsicCurvatures}
\end{equation}
Although the formulas are useful for computing curvature and torsion in practice, the geometrical interpretation of these parameters is somewhat hidden in these expressions. 

\subsection{Local representation of the curves and Frenet frames}\label{sec:Frenet}

As the point in the curve lies in $\mathbb{R}^3$, we can define a three-dimensional basis for each point.
The arclength parametrization of the curve $x(t) = X(s(t))$ implies that $\dot{x}(t)=\dot{s}(t)X'(s(t))$, meaning that the tangent vector $T(s)\triangleq {X}'(s)$ is unit length for all $s$ in $\left[0,L\right]$. 
At points where $\|T(s)\| \neq 0$, since $\|T(s)\|^2=1$, the derivative $T'(s)$ is orthogonal to $T(s)$ and thus there exists a unit vector $N(s) \propto T'(s)$. The curvature can be defined as $s\mapsto\kappa(s)=\| T'(s)\|$, which measures %the angle between two neighboring tangent lines, 
how rapidly the curve pulls away from the tangent line.
Adding this Normal vector $N(s)=\frac{1}{\kappa(s)}T'(s)$ together with the bi-normal vector $B(s)=T(s)\times N(s)$ to the tangent vector $T(s)$ defines a local orthonormal basis system in $\mathbb{R}^3$. Viewing the local basis system as a function of $s$ defines a moving frame, known as Frenet frames.
The torsion $s\mapsto\tau(s)$ is the function that satisfies $B'(s)=-\tau(s)N(s)$ for all $s$ in $\left[0,L\right]$, which measures how rapidly the curve pulls away from the osculating plane determined by the tangent vector and the normal vector. Physically, one can view that a curve can be obtained from a straight line by bending (curvature) and twisting (torsion) \citep{Carmo1976}.
%In short, the Frenet frames provides a geometric curve representation, indexed by functional parameters that have physical interpretations.

\subsection{Geometry of the curve and elastic shape analysis}  \label{sec:elastic}

Instead of treating the shape function $X$ directly, elastic shape analysis \citep{Srivastava2011, SrivastavaKlassen2016} treats curves $x$ as a shape object, which are then compared with a geodesic distance between them defined through optimal deformations. A popular transformation is based on the square root velocity function (SRVF), defined for each curve $x(t)=X(s(t))$ as
\[
q_{x}(t)=\frac{\dot{x}(t)}{\sqrt{\left\Vert \dot{x}(t)\right\Vert }} \,. %=\sqrt{\dot{s}(t)}T\left(s(t)\right) \,.
\]
This can be viewed as a representation of the shape of the curve on a manifold. The distance between two curves is then defined as the $L_2$ distance between $q_x$ and is parametrisation-independent. 

%\subsubsection{Elastic shape framework: preshape space}
The SRVF transformation $F:x\mapsto \dot{x}(t)/\sqrt{\left\Vert \dot{x}(t)\right\Vert}$ helps defining a pre-shape space that is used for characterizing the underlying shape of a given function. The pre-shape space for unit length open curves is $\mathcal{C}^{O}=\left\{ q\in L^{2}(\left[0,T\right],\mathbb{R}^{p})\right\} $
and is simply the hypersphere of $L^{2}(\left[0,1\right],\mathbb{R}^{p})$. %meaning that it is a Hilbert submanifold of $L^{2}$.  
The framework is better suited to explain the variation of curves in the presence of warping. Assume that two curves $x_0, x_1$ are similar in the sense that $x_1 \approx x_0 \circ h$ for a time warping function $h:[0,T]\rightarrow [0, T]$. In order to align the curves $x_0, x_1$ with SRVF, we solve the following minimisation problem that defines at the same time a geodesic distance:
\begin{equation}
d_{srvf}(x_{0},x_{1})=\inf_{O\in SO(3), h\in H_T}\int_{0}^{T}\left\Vert q_{0}(t)- O\sqrt{\dot{h}(t)}q_{1}(h(t)) \right\Vert^{2}dt \,.\label{eq:DistanceSRVF}
\end{equation}
The distance $d_{srvf}$ between two curves $x_{0}$ and $x_{1}$ is invariant to translation, rotation and re-parametrisation.
In the case of multiple curves, the SRVF mean $\tilde{\mu}_{srvf}$ is defined as a Fr\'echet mean that minimizes the average geodesic distance:
\begin{equation} \label{eq:SRVFmean}
\tilde\mu_{srvf} = \arg\min_{\mu} \sum_{i=1}^n d_{srvf}(x_i, \mu) \,.
\end{equation}
An iterative algorithm is used to solve the optimization problem. 
%The iterative steps can be expressed, with a slight abuse of notation, as
%\begin{itemize}
%    \item Given $\mu$, estimate an optimal warping for each $i: \tilde h_i = \arg\min_{h} d_{srvf}(\mu, x_i \circ h)$.
%    \item Update $\mu: \tilde\mu = \arg\min_{\mu} \sum_{i=1}^n d_{srvf}(\mu, x_i \circ \tilde h_i))$. 
%\end{itemize}

%Denote the $L^2$ distance involved in $d_{srvf}$ (\ref{eq:DistanceSRVF}) by
%\begin{equation}
%D_{srvf}(x_0, x_1) =  \int_{0}^{T}\left\Vert q_{0}(t)- O\sqrt{\dot{h}(t)}q_{1}(h(t)) \right\Vert ^{2}dt 
%\end{equation}

As the SRVF representation depends on the first derivative, it reflects some geometry of the curve. In fact, we can express $q_x(t) = \sqrt{\dot{s}(t)}T\left(s(t)\right)$. The relation between $T$ and $\kappa$ as explained in section \ref{sec:Frenet} implies that this representation implicitly depends on the curvature and, to less extent, on torsion. \citet{BrunelPark2019} extend this approach by incorporating the Frenet frames directly in the representation. 
Nevertheless, the dependence on the parameters is still implicit in this framework. 

\subsection{Illustration of interplay between curves and geometry} \label{sec:comparison}

\begin{figure}[!h]
    \centering
    \hspace*{-1cm}\begin{tabular}{ccc}
    \includegraphics[width=0.28\textwidth]{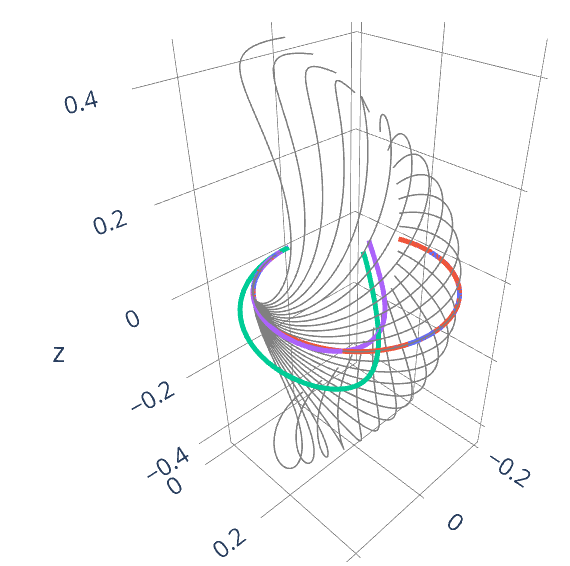} &
    \includegraphics[width=0.35\textwidth]{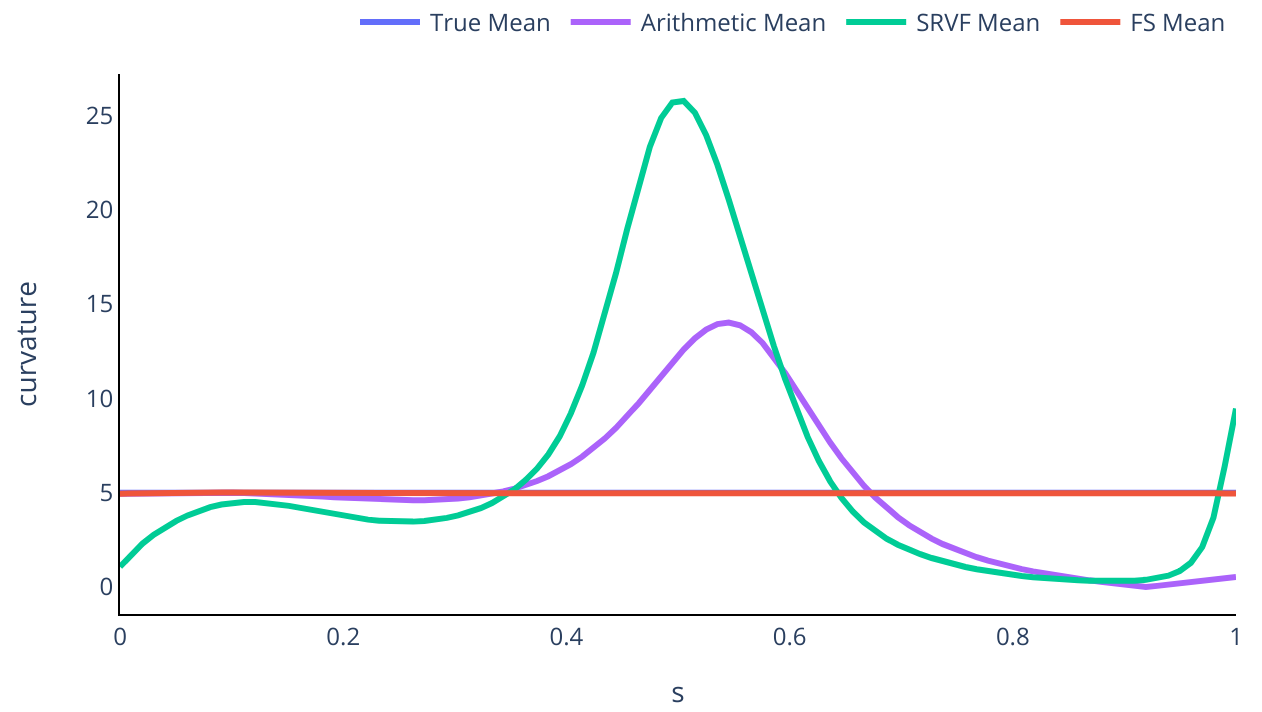} &
    \includegraphics[width=0.35\textwidth]{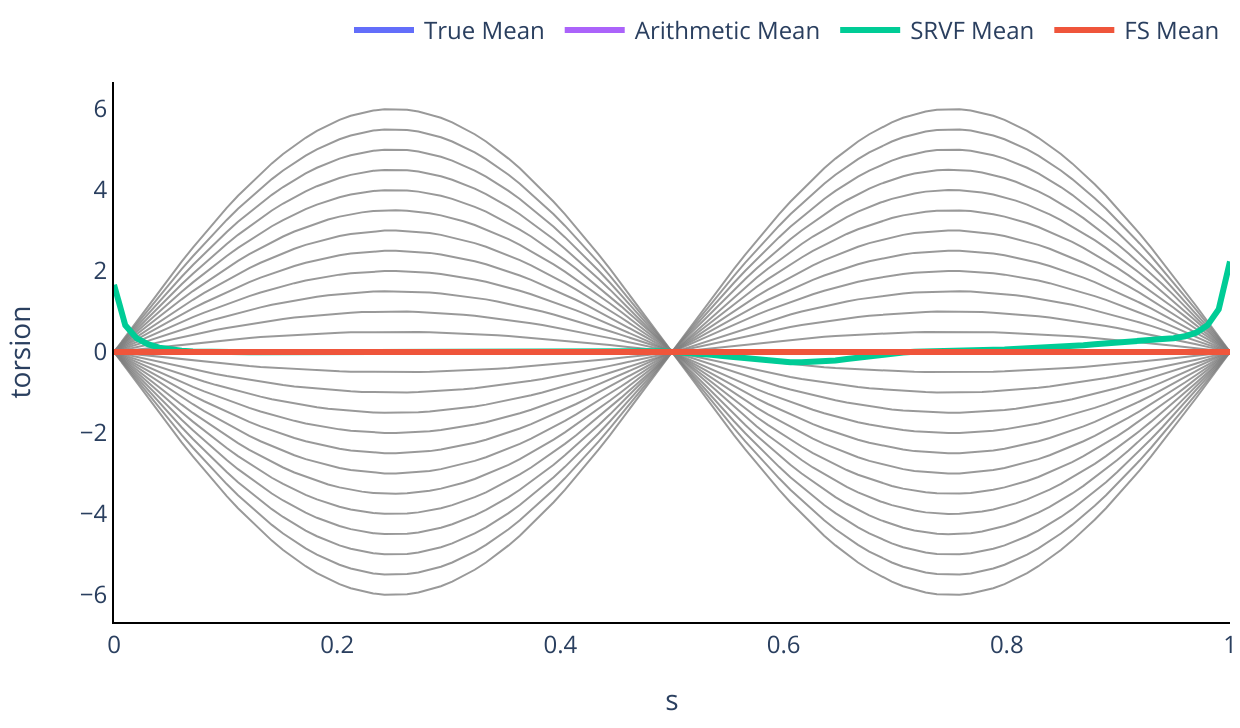} \\
    & (a) Variability in the torsion. & \\
    \includegraphics[width=0.28\textwidth]{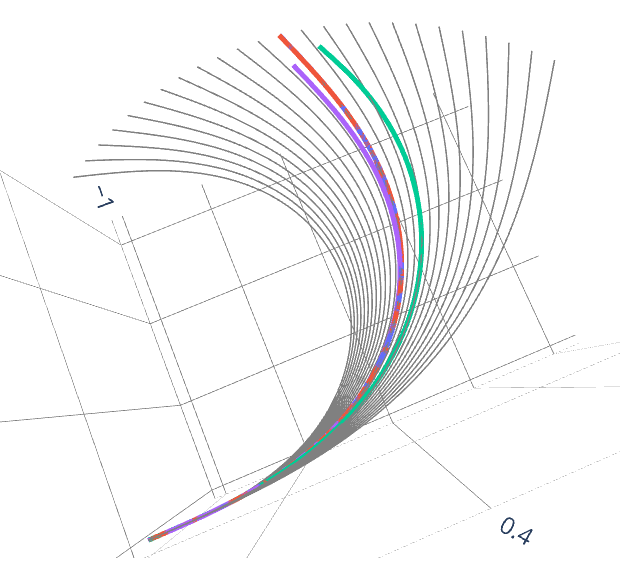} &
    \includegraphics[width=0.35\textwidth]{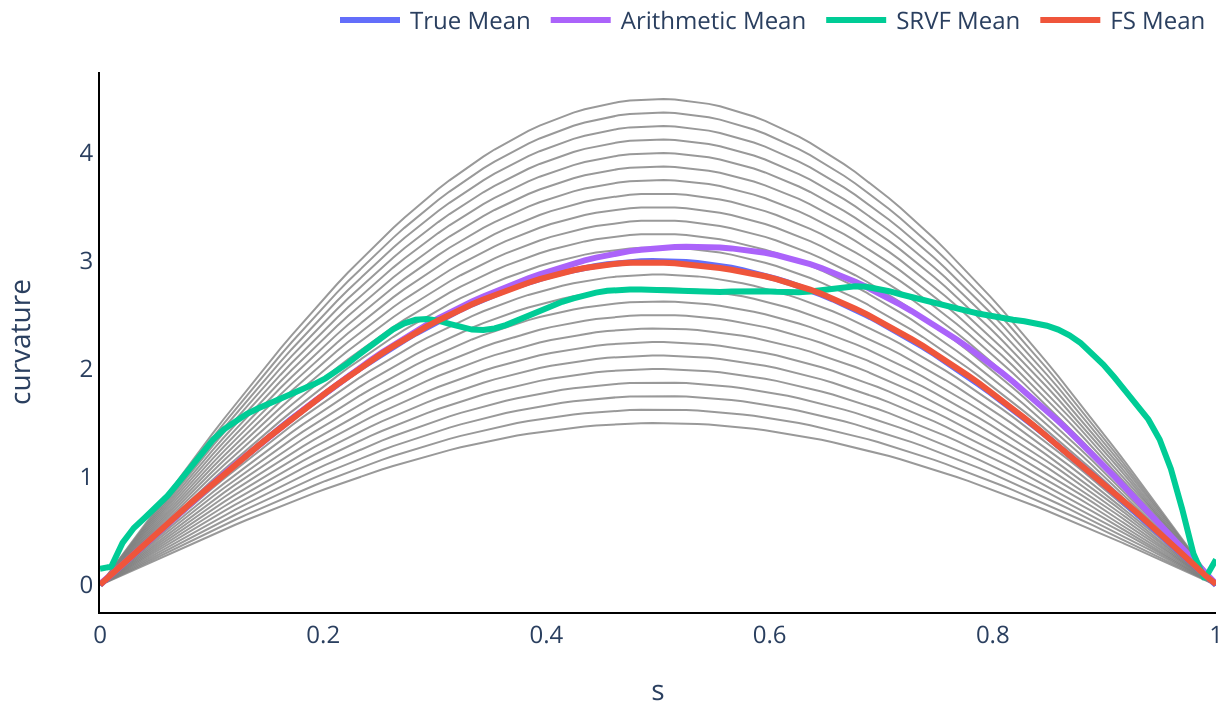} &
    \includegraphics[width=0.35\textwidth]{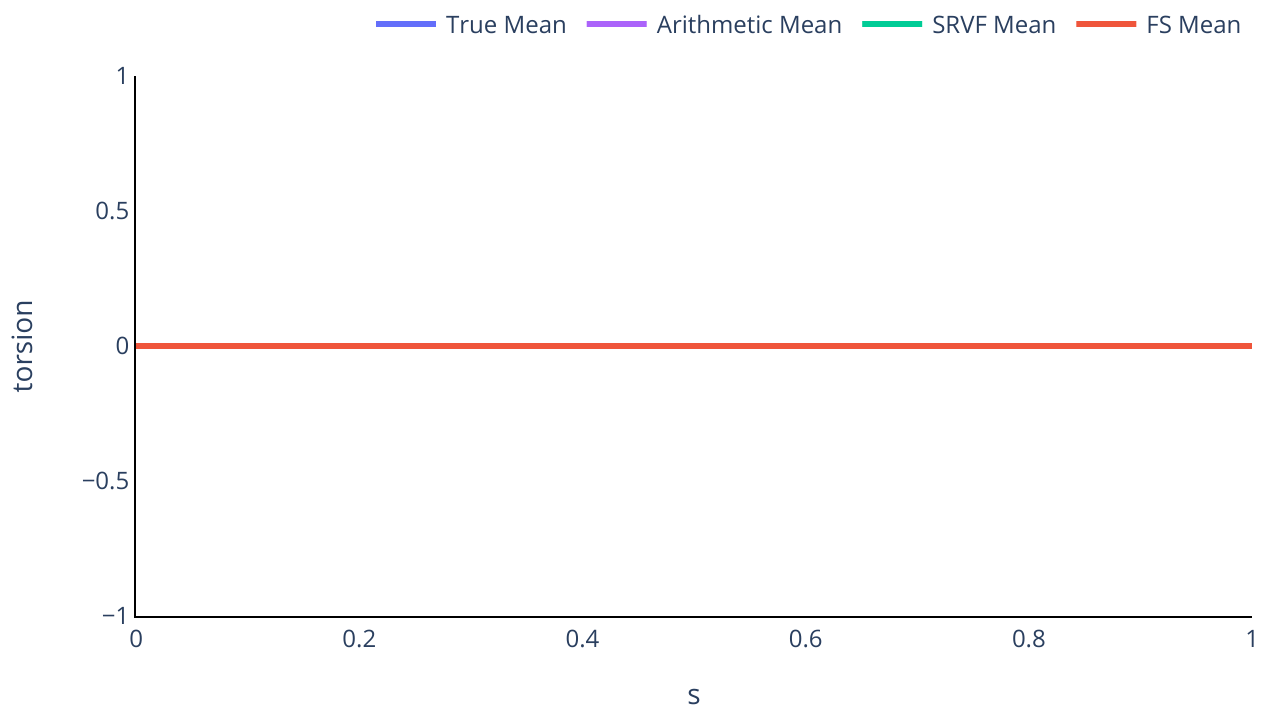} \\
    & (b) Variability in the curvature. &
    \end{tabular}
    \caption{Set of Euclidean curves (first column), curvatures (second column), torsions (third column) overlayed with means: true mean in blue, elastic mean by SRVF in green, arithmetic mean in purple and the proposed mean by Frenet-Serret method in red.}
    \label{fig:Comp_ex}
\end{figure}

To appreciate the significance of curvature and torsion in the representation of curves, we illustrate the interaction between curves and geometry by a simplified example.
We consider two set of $25$ Euclidean curves, the first one with some variability only in the torsion and the second one only in the curvature, as shown in Figure~\ref{fig:Comp_ex}. 
These curves satisfy the geometric curve representation with, 
for the first case, a constant curvature equal to $5$ for all curves and a torsion $\tau_i(s) = - a_i 3 \sin(2 \pi s), i=1,...,25$ with $a_i$ equally spaced from $-1$ to $1$, for the second one, a constant torsion equal to $0$ for all curves and a curvature $\kappa_i(s) = - a_i |3 \sin(\pi s)|, i=1,...,25$ with $a_i$ equally spaced from $-1$ to $1$. 

% !!! without subfigure !!!

% !!! using subfigure !!!
%\begin{figure}[!h]
%    \centering
%    \begin{subfigure}[b]{1\textwidth}
%    \includegraphics[width=0.28\textwidth]{Figures/Comparison_example/Without_smoothQ/TorsVar_Means_grey.png}
%    \includegraphics[width=0.35\textwidth]{Figures/Comparison_example/Without_smoothQ/TorsVar_curv_grey.png} 
%    \includegraphics[width=0.35\textwidth]{Figures/Comparison_example/Without_smoothQ/TorsVar_tors_grey.png}
%    \subcaption[]{Variability in the torsion.}
%    \label{fig:Comp_ex_tors}
%    \end{subfigure} \\
%    \begin{subfigure}[b]{1\textwidth}
%    \includegraphics[width=0.28\textwidth]{Figures/Comparison_example/Without_smoothQ/CurvVar_Means_grey.png}
%    \includegraphics[width=0.35\textwidth]{Figures/Comparison_example/Without_smoothQ/CurvVar_curv_grey.png} 
%    \includegraphics[width=0.35\textwidth]{Figures/Comparison_example/Without_smoothQ/CurvVar_tors_grey.png}
%    \subcaption[]{Variability in the curvature.}
%    \label{fig:Comp_ex_curv}
%    \end{subfigure}
%    \caption{Set of Euclidean curves (first column), curvatures (second column), torsions (third column) overlayed with means: true mean in blue, elastic mean by SRVF in green, arithmetic mean in purple and the proposed mean by Frenet-Serret method in red.}
%    \label{fig:Comp_ex}
%\end{figure}

We compute the mean Euclidean curve by three different methods: the elastic mean by SRVF method (using the implementation in the package \texttt{fdasrsf} \cite{fdasrsf}, the arithmetic mean, and the proposed mean by "Frenet-Serret method", introduced in section ~\ref{sec:FSmean}. The results are shown in the first column of Figure~\ref{fig:Comp_ex}. For two standard methods, there is no connection between the mean and the parameters so we compute the curvature and torsion of each of the means by extrinsic formulas (\ref{eq:ExtrinsicCurvatures}) with estimated derivatives by local polynomial regression. The estimates are shown in the two last columns of Figure~\ref{fig:Comp_ex}. 
It can be seen in the first case that the curvatures of the elastic and arithmetic means are no longer constant, but have a spike between 3 and 5 times larger. 
These two means do not respect the curvature and torsion of the different curves from which they are calculated. The idea of our method is to define a mean that respects the geometry of the curves in all cases. Indeed, the mean calculated by our "Frenet-Serret" method has a constant curvature equals to 5 and a zero torsion.  
Although this example is contrived to demonstrate our motivation, this type of variability between curves can also be found in real data examples, presented in section \ref{sec:FLY}.

%%%%%
\section{Characterization of a Frenet-Serret mean shape \label{sec:FSmean}}

We develop a new framework to characterize a mean shape with an explicit parametrization of the curves linked to the geometry of the curves.
Let us consider $N$ curves in $\mathbb{R}^{3}$ defined as functions $\mathcal{S}=\left\{ x_{1},\dots,x_{N}\right\}$ from $\left[0,T\right]$ to $\mathbb{R}^{3}$. 
%Our idea is to define a mean shape for the population of curves $\mathcal{S}$, through a reference pattern for curves in $\mathcal{C}_{1}$, independently of the variations in translations, rotations and scalings.
For regular curves $x_i$, the shape function $X_i$ is identified with the arclength parameterized curves as $x_i(t)=X_i(s_i(t)), i=1, \ldots, N$.

\subsection{Sources of variation of the curves} \label{sec:timewarping}

The arclength parametrization should not be confused with the standard representation of time warping or phase variation in the functional data. Suppose that $x_i$ is given as $x_i(t) = x(h_i(t))$, where $h_i \in H_T = \{h: [0, T] \rightarrow [0, T]  | h(0)=0, h(T)=T, h^\prime \geq 0\}$ are warping functions. As $\dot{x_i}(t) = {\dot x}(h_i(t))\dot h_i(t)$, by change of variables, the corresponding arclength can be expressed as
\[
s_i(t) = \int_0^t \|{\dot x}(h_i(t)){\dot h_i}(t)\|\,dt = \int_0^{h_i(t)}\|{\dot x}(u)\|\,du = s(h_i(t)) \,.
\]
It follows that  $x_i(t) = x(h_i(t)) = X(s(h_i(t)) = X(s_i(t))$, that is, the \emph{shape} of the curve is preserved under time warping. 
For univariate functional data, phase variation expressed as time warping functions is often confounded with shape variation. % as we introduce here. 

In this work, we distinguish between phase ($h_i$) and shape variation (${X}_i$). As seen earlier, elastic shape analysis is adapted to phase variation.
We explicitly model both types of variations in the spirit of functional data analysis and treat two cases separately 
\begin{equation} \label{eq:xModel}
M_1: x_i(t) = X_i(s_i(t))\,,\qquad M_2: x_i(t) = X_i(s_i(h_i(t))) \,.
\end{equation}
As the latter can be viewed as an extension of the former, we first develop our characterization of mean shape for $M_1$. 
An extension to $M_2$ is presented in section \ref{sec:FrenetMean-warping}.

%under the shape variation of the form 
%\[
%\theta_i(s) = \gamma_i^\prime(s)\theta_0(\gamma_i(s)) \,.\] 

\subsection{Shape function and its equivalent class}\label{sec:FrenetSerretODE-and-Equivalence}

We have seen in section \ref{sec:Frenet} that associated with the shape $X$ is the Frenet frames $T, N, B$, which gives a geometric curve representation. 
In order to link the geometric features of the curve contained in the curvature and torsion to the shape function $X$ of the curve, we first note that the vectors $s\mapsto T(s),\,N(s),\,B(s)$ are tightly related through Frenet-Serret ODE
\begin{equation}
\left\{ \begin{array}{ll}
T'(s)= & \kappa(s)N(s)\\
N'(s)= & -\kappa(s)T(s)+\tau(s)B(s)\\
B'(s)= & -\tau(s)N(s)
\end{array}\right.\label{eq:FS-ode-vector}
\end{equation}
with an initial condition $T(0),N(0),B(0)$. In other words, the moving frame defines a curve $s\mapsto Q(s)=\left[T(s)\vert N(s)\vert B(s)\right]$ in the group of special orthogonal matrices $SO(3)$ where $SO(p) = \{Y \mbox{ is a } p\times p \mbox{ matrix }  \vert Y^\top Y = I_p, \mbox{det}(Y) = 1\}$. 
As $SO(3)$ is a Lie group with a manifold structure, the Frenet-Serret ODE can be seen as an ODE defined in the Lie group with
\begin{eqnarray}
Q^\prime(s) & = & Q(s)A(s)\label{eq:FS-ode-matrix}
\end{eqnarray}
where 
\begin{equation}
A(s)=\left[\begin{array}{ccc}
0 & -\kappa(s) & 0\\
\kappa(s) & 0 & -\tau(s)\\
0 & \tau(s) & 0
\end{array}\right]\,,\label{eq:matrixA}
\end{equation} 
and $A^\top = -A$ so $A$ is skew-symmetric. 
We shall denote by $\theta$ the functional parameters $(\kappa, \tau)$ with the set of admissible parameters by $\mathcal{H} = \{\theta = (\kappa,\tau), \kappa>0,  \kappa, \tau \in C^2 \}$, and by $A_\theta$ the corresponding skew-symmetric matrix.
We call the solutions $s\mapsto Q(s)$ of the Frenet-Serret equations the Frenet paths, and the set of Frenet paths is denoted by 
\[
\mathcal{F}=\left\{ s\mapsto Q_{\theta}(s)\vert Q^{\prime}(s)=Q(s)A_{\theta}(s),\:s\in\left[0,1\right],\,Q(0)\in SO(3),\,\theta\in\mathcal{H}\right\} \,.
\]
%The set of admissible parameters is defined as $\mathcal{H} = \{\theta = (\kappa,\tau), \kappa>0, (\kappa, \tau) \mbox{ are $r$ times continuously differentiable} \}.$
Among the set of all Frenet paths, we pay a particular attention to 
the subset of Frenet paths with initial condition equal to the
identity matrix $I_{3}$, $\mathcal{F}_{0}=\left\{ Q\in\mathcal{F}\vert Q(0)=I_{3}\in SO(3)\right\}$.
Denote the set of arclength-parametrized regular curves of length 1 by $\mathcal{C}_{1}$. As any regular curve $X$ can be recovered by integrating its tangent $X'(s)=T_{\theta}(s)$,
we have 
\begin{equation}
\mathcal{C}_{1}=\left\{ s\mapsto {X}(s)=X_{0}+Q_{0}\int_{0}^{s}T_{\theta}(u)du\:\vert\:Q_{\theta}\in\mathcal{F}_{0},\,X_{0}\in\mathbb{R}^{3},\,Q_{0}\in SO(3)\right\} \label{eq:Space_Regular}
\end{equation}
indexed by the parameters $\left(X_{0},Q_{0},\theta\right)$.
This parametrization is known to be one-to-one: for any curve $X_{1}$ and $X_{2}$ having the same curvature and torsion, there exists a vector $a$ and a rotation $R\in SO(3)$ such that ${X}_{1}=a+R{X}_{2}$. 
%From that parametrization, it is clear that the functions $s\mapsto\theta(s)$ or $s\mapsto Q_{\theta}(s)$ represents the geometrical content of any regular curve $\mathbf{X}$.
If the Frenet path for ${X}_{2}$ has an initial condition
equal to $I_{3}$, the rotation matrix $R$ is exactly the initial
condition of the Frenet path associated with ${X}_{1}$. For
this reason, the space $\mathcal{F}_{0}$ can be naturally considered
as the shape space. %}
The functions $s\mapsto\theta(s)$ or $s\mapsto Q_{\theta}(s)$ represent the geometrical content of any regular curve ${X}$.

For regular curves in $\mathbb{R}^{p}$ for $p>3$, the same moving
frame in $SO(p)$ can be defined, in terms of a skew-symmetric matrix
similar to (\ref{eq:matrixA}), and the so-called generalized
curvatures $\kappa_{1},\kappa_{2},\dots,\kappa_{p-1}$ \citep{kuhnel2015differential}. 
%From the construction of the Frenet-Serret framework, the set of admissible curvatures is defined as $\mathcal{H}^{p}=\left\{ \theta=\left(\kappa_{1},\kappa_{2},\dots,\kappa_{p-1}\right),\,\kappa_{i}\in H^{1}(0,1),\,\kappa_{i}>0,\,i=1,\dots,p-1\right\}$, where $H^1$ is the Sobolev space of order 1. 
%, $H^1 = \{f: f^{(r)} \mbox{abs. cont.} r=0, 1, f^{(\prime\prime} \in L_2(0,1)}$, 

\subsection{Effect of scaling} \label{sec:scaling}

If we want to consider invariance with respect to rescaling, it suffices to rescale the curves of different length $L_i$ (and arclength $s_i$) to the same length equal to $1$. %For invariance with respect to scaling,  we define the equation with the scaled curves to the unit length.
Rescaling does change the geometry only through a scaling factor, i.e the matrix $s\mapsto A(s)$ in the ODE (\ref{eq:FS-ode-matrix}) is also renormalized and the rescaled curves
$\frac{1}{L}{X}(s)$ have new curvilinear arclength
$\tilde s=s/L$ and the rescaled Frenet paths are $\tilde{Q}(\tilde s)=Q(\tilde sL)$.
The rescaled Frenet-Serret ODE, defined on $\left[0,1\right]$ is
$\tilde{Q}^{'}(\tilde s)=\tilde{Q}(\tilde s)\tilde{A}(\tilde s)$ with $\tilde{A}(\tilde s)=LA(\tilde sL)$,  implying that rescaling a curve by $1/L$, multiplies its curvature and torsion by $L$. 

From now on, we define the equation with the scaled curves to the unit length.

\subsection{Problem formulation}

Recall that the curves can be expressed as
\[
x_i(t) = X_i(s_i(t))\,,\quad s_i(t)=\int_{0}^{t}\Vert \dot{x}_i(u)\Vert\,du \,, \quad i=1,\ldots,N\,.
\]
As seen in section \ref{sec:FrenetSerretODE-and-Equivalence}, the shape function $X_i$ is associated with an ODE parametrized with respect to functional parameter $\theta_i$: 
\[
X_i(s) = X_{i0} + Q_{i0}\int_0^s T_{\theta_i}(u)\,du\,,\quad {Q}_i'(s) = Q_i(s)A_{\theta_i}(s) \,.
\]
Consequently, we identify the shapes with the Frenet paths 
$\boldsymbol{Q}=\left\{ Q_{1},\dots,Q_{N}\right\} $, or equivalently with the set of curvatures and torsions $\boldsymbol{\theta}=\left\{ \theta_{1},\dots,\theta_{N}\right\}$.
Our aim is then to derive a mean parameter $\bar{\theta}$ (and mean Frenet path $\bar{Q}$) for $\mathcal{S}$ as a measure of centrality that corresponds to the mean shape defined as
\[
\bar X(s) = \bar X_0 + \bar Q_0\int_0^s T_{\bar\theta}(u)\,du\,,\quad {\bar Q}'(s) = \bar Q(s)A_{\bar \theta}(s) \,,
\]
which is independent of the variations in translations, rotations and scalings.
Our parametrization of curves in (\ref{eq:Space_Regular}) shows that the quotient space of arclength parametrized curves (under the group action of Euclidean motions) is exactly the space of Frenet paths.
Hence, it is sufficient to work with a population of Frenet paths to define a mean shape.

Contrary to the elastic shape analysis, our focus is not on defining a proper metric on the shape space to define a Fr\'echet mean. 
%This would lead to the elastic shape analysis with the Frenet frames discussed in section \ref{sec:elastic}, with the same limitations.
We are interested in developing a statistical characterization of a mean shape that enables us to identify the mean parameter. %by exploiting the properties of the Frenet paths as the solution to an ODE.
We do not assume the existence of a generative model for the mean shape or the mean parameter in relation to $Q_i$ or $\theta_i$ but directly exploits the characteristic features of the Frenet paths, as the solution of the ODEs, and consider the ODE as a model constraint.

%We develop the characterization of these quantities below. %Hence, the main question is how to define $\bar Q$ and $\bar\theta$.

%In order to define a mean shape based on the Frenet-Serret representation, we first study the characteristic features of Frenet Paths.

%%%
\subsection{Mean shape and mean vector field \label{sec:Mean-shape-ODE}}

A fundamental concept for solving an ODE is the flow over time $t$, denoted by $\phi(t, \cdot)$ \citep{Hairer2006}. It is the mapping that, to each point $Q \in SO(p)$, associates the value of the Frenet paths $Q(t)$ at time $t$ of the solution with initial value $Q(0) = Q$. That is, $\phi(t, Q) = Q(t)$ if $Q(0) = Q$ and $t$ represents the elapsed time. To express the dependence on the initial time, we extend the definition of the flow as $\phi(t, s, Q) = Q(s+t)$ if $Q(s) = Q$ so $\phi(t-s, s, Q) = Q(t)$.
The essential property of the flow is the group property, i.e for all $s,u,t\in\left[0,1\right]$ and $Q\in SO(p)$, $\phi\left(t-s,s,Q\right)=\phi\left(t-u,u,\phi\left(u-s,s,Q\right)\right)$. 
This allows us to express any localized solution coherently to the global solution. As the Frenet path is indexed by $\theta$, the corresponding flow is written as $\phi_\theta$.

We see that the geometrical features $\theta_{i}$ define the vector field $Q\mapsto QA_{\theta_{i}}(s)$, and that the observable features such as tangent, normal or binormal vectors are in fact the corresponding flows $\phi_{\theta_i}$. 
These observations lead us to defining the mean shape
as the mean of the vector fields $Q\mapsto QA_{\theta_{i}}(s)$. We define then the mean vector field as the vector field defined on $SO(p)$ such that the solution paths are close to the individual Frenet paths $Q_{i},\:i=1,\dots,N$. In other words, the
mean vector field corresponds to the flow that provides {\it a best} approximation to all the individual flows. 

A noticeable feature of our formulation is that we do not use the infinitesimal characterization of the differential equation based on the derivative. We use instead the group property of the flow that can be interpreted as a \emph{self-prediction property}: if $s\mapsto Q(s)$ is a solution to equation (\ref{eq:FS-ode-matrix}), then for all $t,\,s\in\left[0,1\right]$ such that $\left|t-s\right|\leq1$, we have 
\begin{equation}
Q(t)=\phi_{\theta}\left(t-s,s,Q(s)\right).\label{eq:selfprediction}
\end{equation}
Otherwise, the curve $s\mapsto Q(s)$ is a solution to $Q'=QA_{\theta}$
if and only if 
\begin{equation} \label{eq:selfprediction2}
\int_{0}^{1}\int_{0}^{1}d\left(Q(t),\phi_{\theta}\left(t-s,s,Q(s)\right)\right)dsdt=0,
\end{equation}
where $d(\cdot,\cdot)$ is a distance defined on $SO(p)$. As $d$ is non-negative, it holds also with $d^2$ in (\ref{eq:selfprediction2}).
%In the case of multiple solutions $Q_{1},\dots,Q_{N}$, it holds that
%\[
%\sum_{i=1}^N \int_{0}^{1}\int_{0}^{1}d\left(Q_i(t),\phi_{\theta_i}\left(t-s,s,Q_i(s)\right)\right)^2 = 0\,.
%\]
As we want to replace $\theta_i$ by a common $\bar\theta$, we require that the mean flow $\phi_{\bar\theta}$ should minimize the self-prediction errors for all the trajectories simultaneously. The individual error is measured by
\begin{equation} \label{eq:selfpredCrt}
\mathcal{V}(Q_i, \phi_\theta) = \int_{0}^{1}\int_{0}^{1}d\left(Q_i(t),\phi_{\theta}\left(t-s,s,Q_i(s)\right)\right)^2 dsdt \,.
\end{equation}
\begin{definition} \label{def:MeanShape1}
Let $Q_i \in SO(p), i=1,\ldots, N$ be the independent and identically distributed random Frenet paths with the same distribution as $Q$, associated with parameters $\theta_i\in\mathcal{H}$ satisfying $Q_i' = Q_iA_{\theta_i}$. %where $\theta_i$ are independent and identically distributed random sample from an underlying population. 
The mean parameter for the Frenet path $Q$ is defined as
\[
\bar\theta = \arg\min_{\theta\in \mathcal{H}} E \{\mathcal{V}(Q, \phi_\theta)\} \,.
\]
%\[
%\bar\theta = \arg\min_{\theta\in \mathcal{H}} \frac{1}{N}\sum_{i=1}^N \mathcal{V}(Q_i, \phi_\theta) \,.
%\]
\end{definition}
%In this work, the prediction error is measured with the geodesic distance in $SO(p)$. 

%%
\subsection{Estimation of mean parameter} \label{sec:Stat-crt}

Suppose that we have a sample of Frenet paths $\{Q_1,\ldots, Q_N\}$ with the corresponding parameters $\theta_1,\ldots,\theta_N$. 
%In addition, identifying the flow requires solving the ODE on $SO(3)$. 
We develop an empirical criterion to construct an estimator. 
%To account for estimation of the functional parameter, we will consider a regularized version here.
The essential ingredients of our definition of the mean based on the self-prediction criterion (\ref{eq:selfpredCrt}) are the distance function $d$ and the representation of the flow $\phi_\theta$. The choice of these need to be adapted to the underlying sample space. 

%In particular, it is convenient to consider the Frenet-Serret equation as an ODE on the Lie group $SO(p)$. 
Due to the orthogonality constraint, the Frenet differential equation is not defined on the Euclidean space but on the special Lie group $SO(p)$. 
Ensuring the orthogonality constraint requires a special treatment in developing a numerical algorithm to solve an ODE and also in tackling a parameter estimation problem in ODE, as numerical errors can accumulate and induce an uncontrolled bias. 
The extension of the theory of ODEs from Euclidean space to Lie groups or manifolds is well developed \citep{Hairer2006}. % and it usually relies on the property of smooth space and its ability to define a tangent space. 
In particular, the rotation group $SO(p)$ is a Lie Group that is also a differentiable manifold, with many remarkable properties that are essential in tackling the numerical problems \citep{Absil2010}. 

We first highlight some useful features of the sample space as $SO(p)$, which allows us to define a geodesic distance $d$ and develop a workable representation of the flow $\phi_\theta$. Based on these, we develop an empirical criterion in the spirit of nonparametric function estimation problem. 

%In order to define a mean shape based on the Frenet-Serret representation, 
%Exploiting the geometric representation, we propose a method to identify the mean shape through the notion of a mean vector field. 
%We first review basic properties of $SO(p)$ and introduce some important concepts in view of statistical shape analysis that provide the basis of our formulation. In particular, we provide a link between the skew-symmetric matrix $s\mapsto A_{\theta}(s)$ (or the curvature $\theta$) and the Frenet path $s\mapsto Q(s)$ defined by the corresponding Frenet-Serret ODE.
%It is convenient to consider the Frenet-Serret equation as an ODE on the Lie group $SO(p)$. 

\subsubsection{Solving ODE on $SO(p)$}

Typically, $SO(p)$ is considered as a submanifold of the Euclidean space $\mathbb{R}^{p\times p}$, with the usual inner product $\left\langle M,N\right\rangle =\mbox{Tr}\left(M^{\top}N\right)$
(and the associated Frobenius norm).
The Tangent Space at point $M$ to $SO(p)$ is the vector space 
\[
T_{M}SO(p)=\left\{ MU\vert U^{\top}=-U\right\}\,,
\]
usually identified with the set of skew-symmetric matrices ($U$). In particular, the Tangent Space at the identity $I_{p}$ is called the Lie algebra of the Lie group, denoted by $\mathfrak{so}(p)$.

A fundamental tool for the analysis of ODE and flows on Lie groups
is the Exponential map, $\mbox{Exp}_{M}$, at point $M$, which relates the tangent space to the manifold. The Exponential map $\mbox{Exp}_{M}\,:\,T_{M}SO(p)\longrightarrow SO(p)$
is such that $\mbox{Exp}_{M}(U)=\gamma(1;M,U)$, where $\gamma$ is
the unique geodesic $s\mapsto\gamma(s,M,U)$ such that $\gamma(0;M,U)=M$
and $\gamma'(0;M,U)=U$. Conversely, if we have a given root
$M$ and a target point $N$, the logarithmic map returns a tangent
vector at $M$, pointing toward $N$, of length $dist(M,N)$. Hence,
the logarithmic map $\mbox{Log}_{M}\,:\,SO(p)\longrightarrow T_{M}SO(p)$ at $M$ is $\mbox{Log}_{M}(N)=V$ such that $\mbox{Exp}_{M}(V)=N$ and $\| \mbox{Log}_{M}(N)\| =dist(M,N)$. 

Fortunately, if we consider a matrix Lie group, the exponential
and logarithmic maps can be expressed simply with the classical matrix exponential $\exp(A)=\sum_{k\geq0}\frac{A^{k}}{k!}$ and matrix logarithm, see \citet{Higham2008}. In particular, we have %for $M, N\in SO(p)$ and $U\in \mathfrak{so}(p)$, 
\begin{equation} \label{eq:SOpExpLog}
\mbox{Exp}_{M}(U)=M\exp(U)\,,\quad \mbox{Log}_{M}(N)=\log(M^{\top}N) \,.
\end{equation} 
As a consequence, the geodesic distance $dist(M,N)=\| \log(M^{\top}N)\|_{F}$ has a closed form expression that is amenable to computation. 
%We evaluate  the criterion (\ref{eq:selfpredCrt}) in terms of the geodesic distance.
Numerous efficient algorithms exist for computing the exponential of a matrix; the case of $p=3$ is remarkable, as in that case the exponential and logarithm have a closed-form expression.
%given by the so-called Rodrigues formula. 
We will use in our applications these formulas to derive our fast algorithms. 

%\subsubsection{Solving Ordinary Differential equations on the $SO(p)$}
Now we want to express the flow $\phi_\theta$ of the Frenet ODE on $SO(p)$. An ODE is defined as a function $F\::\:SO(p)\longrightarrow T_{M}SO(p)$, such that $\dot{Y}(t)=F\left(t,Y(t)\right)$. In the case of the Frenet-Serret equation, the vector field is time-varying but relatively simple.
In light of the relation (\ref{eq:SOpExpLog}), a fruitful approach to solving a differential equation $\dot{Y}=YA(t)$ with $Y(0)=Y_{0}$ in a Lie group is to look for a solution of the form $Y(t)=Y_{0}\exp\left(\Omega(t)\right)$, which defines the flow. This implies that the function $t\mapsto\Omega(t)$ is defined in $\mathfrak{so}(p)$ and is known to admit the so-called Magnus expansion (chapter IV.7 in \citet{Hairer2006}, \citet{Iserles2000})
%and is itself a solution of the \emph{dexpinv} differential equation (chapter IV.7 in \citet{Hairer2006}, \citet{Iserles2000}).
%Consequently, the function admits the so-called Magnus expansion
{\small
\begin{equation}
\hspace*{-0.1cm}\Omega(t)=\int_{0}^{t}  A(s)ds-\frac{1}{2}\int_{0}^{t}\petit\left[\int_{0}^{\tau}\petit A(s)ds,A(\tau)\right]\petit d\tau +\frac{1}{4}\int_{0}^{t}\petit\left[\int_{0}^{\tau} \left[\int_{0}^{\sigma}\petit A(\mu)d\mu,A(\sigma)\right]\petit d\sigma,A(\tau)\right]\petit d\tau+\dots\,,\label{eq:MagnusExpansion}
\end{equation}
}
which can be used to derive efficient integration methods. 
Additional properties of the matrix exponential are summarized in the supplementary.

%Utilizing the Magnus expansion, we obtain an exponential form for the flow of
%the ODE (\ref{eq:FS-ode-matrix}). 
%The dependence of the solution to the ODE on the initial condition can be expressed in terms of the flow $\phi_{\theta}$. It is the function such that for all $s$ in $\left[0,1\right]$, $t\mapsto\phi_{\theta}\left(t,s,M\right)$ is the solution of the Frenet-Serret ODE satisfying $Q(s)=M$. 

As explained in section~\ref{sec:Mean-shape-ODE}, the flow can be generalized to represent initial values at arbitrary time $s$. For all $t,s$ in $\left[0,1\right]$ and $\theta$ in $\mathcal{H}$, we define the matrix-valued (in $\mathfrak{so}(p)$) function $\Omega\left(t,s;\theta\right)$ such that the flow can be written as 
\begin{equation}
\phi_{\theta}\,:\,\left(t,s,Q\right)\mapsto Q\exp\left(\Omega(t,s;\theta)\right) \,,\label{eq:ExactODEFlow}
\end{equation}
where $\Omega(t,s)$ is defined by replacing $\int_0^t$ with $\int_s^{s+t}$ in (\ref{eq:MagnusExpansion})
so that $Y(t) = \phi_\theta(t-s, s, Y_0)$ if $Y(s)=Y_0$ and $\Omega(t,s; \theta)$ expresses the dependence on $\theta$.

%The essential property of the flow is the group property, i.e for all $s,u,t\in\left[0,1\right]$ and $Q\in SO(p)$, $\phi_{\theta}\left(t,s,Q\right)=\phi_{\theta}\left(t-u,u,\phi_{\theta}\left(u-s,s,Q\right)\right)$. This allows us to express any localized solution coherently to the global solution.

%In the computation and in the analysis of approximation error of our algorithms, we will use repeatedly the commutator $\left[A,B\right]$ between $A$ and $B$ in the Lie algebra $\mathfrak{so}(p)$ defined as $\left[A,B\right]=AB-BA$. More generally, the commutator between two vector fields measures and computes the degree of non-commutativity between two vector fields. In the case of matrix Lie groups, it boils down to the classical matrix commutator $\left[A,B\right]$, often denoted by $\mbox{ad}_{A}(B)$ (derivative of the adjoint representation). The commutator arises in the Baker-Campbell-Hausdorff (BCH) formula, which is central in the theoretical and computational analysis of functions on Lie Groups as 
%\begin{equation}
%\exp(tA)\exp(tB)=\exp\left(tA+tB+\frac{1}{2}t^{2}\left[A,B\right]+O(t^{2})\right)\label{eq:BCHformula}
%\end{equation}
%for $t$ small enough. 

\subsubsection{Estimation criterion} \label{sec:estCrt}

%In this work, the prediction error is measured with the geodesic distance in $SO(p)$. 
Using the geodesic distance in $SO(p)$, combined with the flow (\ref{eq:ExactODEFlow}), the criterion (\ref{eq:selfpredCrt}) can be expressed as
\[
\mathcal{V}(Q_i, \phi_\theta) = \int_0^1 \int_0^1 \left\Vert\log\left(Q_{i}(t)^{\top}Q_{i}(s)\exp\left(\Omega(t-s,s;\theta)\right)\right)\right\Vert_{F}^{2} \,dsdt \,, \quad i=1, \ldots, N\,.
\]
To allow for variation in the prediction error, we incorporate weights according to the distance to initial values $s$ in evaluating the solution at $t$ and define a weighted criterion:
%{\footnotesize\begin{equation}
%%\breve{\mathcal{I}}_{h,\lambda}(\theta;\boldsymbol{Q})
%\frac{1}{N}\sum_{i=1}^{N}\int_{0}^{1}\int_{0}^{1}K_{h}(t-s)\| \log\left(Q_{i}(t)^{\top}Q_{i}(s)\exp\left(\Omega(t-s,s;\theta)\right)\right)\| _{F}^{2}dsdt+\lambda\int_{0}^{1}\| \theta^{''}(t)\| ^{2}dt \label{eq:SelfPrediction1}
%\end{equation}}
\begin{equation} \label{eq:SelfPrediction1}
\breve{\ell}_{N,h}(\theta) =    \frac{1}{N}\sum_{i=1}^{N}\int_{0}^{1}\int_{0}^{1}K_{h}(t-s)\| \log\left(Q_{i}(t)^{\top}Q_{i}(s)\exp\left(\Omega(t-s,s;\theta)\right)\right)\|_{F}^{2} \,dsdt \,,
\end{equation}
where $K(\cdot)$ is a kernel function with compact support,
e.g. $K(u)=\frac{3}{4}(1-u)^{2}1_{[-1,1]}(u)$ and  $K_{h}(u)=(1/h)K(u/h)$.
%As usual, we denote the scaled kernel by $K_{h}(u)=(1/h)K(u/h)$.
% not necessary here
%, and the absolute moments of $K$ are denoted by $\mu_m(K)=\int_{-1}^{1}\left|x\right|^{m}K(x)dx$. %\sigma_{K}^{m}=\int_{-1}^{1}\left|x\right|^{m}K(x)dx
The kernel $K(\cdot)$ and the bandwidth $h$ define a prediction
horizon for the flow.
In addition, we introduce a smooth regularization for the functional parameter $\theta$ with a penalty term
\begin{equation}\label{eq:Penalty}
\mathcal{P}_\lambda(\theta) = \lambda \|\theta^{\prime\prime}\|_2^2
= \lambda\int_{0}^{1}\| \theta^{''}(t)\|^{2}dt \,,
\end{equation}
and define the empirical criterion as $\breve{\mathcal{I}}_{h,\lambda}(\theta) = \breve{\ell}_{h}(\theta) + \mathcal{P}_\lambda(\theta)$.
\begin{definition}
Let $\{Q_1,\ldots, Q_N\}$ be a sample of Frenet paths with parameters of curvature and torsion $\theta_1,\ldots,\theta_N$. For a fixed $h$ and $\lambda$, the sample mean vector field (or curvature) is defined as the parameter $\theta$ that minimizes the global prediction error $\breve{\mathcal{I}}_{h,\lambda}(\theta)$.
%, that is,
%\[
%\breve{\theta}_{h,\lambda} = \arg\min_{\theta\in\mathcal{H}} \breve{\mathcal{I}}_{h,\lambda}(\theta;\boldsymbol{Q}) \,.
%\]
\end{definition}
Our definition can be viewed as a generalization of the mean in the scale-space view in nonparametric curve estimation \citep{ChaudhuriMarron2000, Wei2018}. The following Proposition shows that the mean vector
field exists for any $h$ and $\lambda$ in great generality, %SO(3)
as long as the sample is bounded in $L^{2}$. 

%The existence is addressed in the next proposition. 
%We show that the mean vector field $\breve{\theta}_{h,\lambda}$ exists under general conditions.
\begin{prop}
\label{ExistenceUniquenessMean} Let $Q_{1},\dots,Q_{N}$ be Frenet
paths with parameters $\boldsymbol{\theta}$, such that for all $i=1,\dots,N$,
$\| \theta_{i}\| _{\infty}\leq C$. There exists
$\breve{\theta}_{h,\lambda}$ in $\mathcal{H}$ such that 
\[
\breve{\theta}_{h,\lambda}\in\arg\min_{\theta\in\mathcal{H}}\breve{\mathcal{I}}_{h,\lambda}(\theta)\,.
\]
\end{prop}
We can also define the mean Frenet Path $\breve{Q}_{h,\lambda}(t)=\exp\big(\Omega(t,0,\breve{\theta}_{h,\lambda})\big)$ and the corresponding mean shape $\breve{{X}}$ obtained by integrating the gradient. 
However, it is rather difficult to compute the corresponding mean or to analyze it. 
Since the expression of $\Omega$ is intractable in general, we further derive a consistent approximations to the flow, by truncating the Magnus expansion, see chapter IV in \citet{Hairer2006}. 
%Recall that the criterion is based on a distance $d$, then our idea is to choose $\tilde\phi_\theta$ such that
%\[
%d(Q_i, \phi_\theta)^2 \leq d(Q_i, \tilde{\phi}_\theta)^2 + d(\tilde{\phi_\theta}, \phi_\theta)^2
%\]
%where $d(\tilde{\phi_\theta}, \phi_\theta)$ is small.
In particular, we use an approximation of order 2, obtained by using a simple quadrature rule with the midpoint and truncating after the first term: $Q_{s+h}=Q_{s}\exp\left(hA_{\theta}\left(s+\frac{h}{2}\right)\right)$, i.e. $\phi_{\theta}(h,s,Q_{s})-Q_{s+h}=O(h^{2})$. The corresponding approximate flow $\tilde{\phi}_{\theta}\left(h,s,Q\right)=QN_{h}(s,\theta)$ can be seen as an Euler-Lie method that possesses several interesting features: it respects the $SO(p)$ constraint, has an explicit and pointwise dependence in $\theta$, and the approximation is uniform on $SO(p)$. 
For this reason, we introduce an approximation, $\mathcal{I}_{h,\lambda}(\theta) = \ell_{N,h}(\theta) + \mathcal{P}_\lambda(\theta)$, to the criterion $\mathcal{\breve{I}}_{h,\lambda}$, valid for small $h$ (S2.1 in the supplementary), where
%\begin{equation*} \label{eq:SelfPrediction2}
%\hspace*{-0.5cm}\ell_{N,h}(\theta; \boldsymbol{Q})
%= \frac{1}{N}\sum_{i=1}^{N}\iint_{0}^{1}K_{h}(t-s) \| \log\left(\exp\left(-(t-s)A_{\theta_{i}}\left(\frac{s+t}{2}\right)\right)\exp\left((t-s)A_{\theta}\left(\frac{s+t}{2}\right)\right)\right)\| _{F}^{2}dsdt  \,.
%\end{equation*}
%\comment{In fact, we only need:}
\begin{equation*} \label{eq:SelfPrediction2}
\hspace*{-0.5cm}\ell_{N,h}(\theta)
= \frac{1}{N}\sum_{i=1}^{N}\int_0^1\int_{0}^{1}K_{h}(t-s) \left\Vert \log\left(Q_i(t)^\top Q_i(s)\exp\left((t-s)A_{\theta}\left(\frac{s+t}{2}\right)\right)\right)\right\Vert _{F}^{2}dsdt  \,.
\end{equation*}
The following proposition shows that, at first approximation, our approach is tractable and can be easily understood in terms of the geometry the curves. 
%{\footnotesize\begin{eqnarray}
%%\mathcal{I}_{h,\lambda}(\theta)
%&&\hspace*{-1cm}\frac{1}{N}\sum_{i=1}^{N}\int_{0}^{1}\int_{0}^{1}K_{h}(t-s) \| \log\left(\exp\left(-(t-s)A_{\theta_{i}}\left(\frac{s+t}{2}\right)\right)\exp\left((t-s)A_{\theta}\left(\frac{s+t}{2}\right)\right)\right)\| _{F}^{2}dsdt \nonumber \\
%&& \quad +\lambda\int_{0}^{1}\| \theta^{''}(t)\| ^{2}dt\label{eq:SelfPrediction2}
%\end{eqnarray}}
%%%%%%%%%%%%%
\begin{prop}\label{ApproxCrt}
Let $Q_{1},\dots,Q_{N}$ be Frenet paths with parameters $\theta_{i},\:i=1,\dots,N$
in $\mathcal{H}$, satisfying $\| \theta_{i}\| _{2}^{2}\leq\frac{\pi}{2}$. 
Then, there exists $B>0$, such that for all $\| \theta\| _{2}\leq B$,
\[
\breve{\mathcal{I}}_{h,\lambda}(\theta)-\mathcal{I}_{h,\lambda}(\theta)=O(h^{3}).
\]
%For small $h$, there exists a minimum 
%\[
%\theta_{h,\lambda}\in\arg\min_{\theta\in\mathcal{H}}\mathcal{I}_{h,\lambda}(\theta) \,,
%\]
%such that $\theta_{h,\lambda}-\breve{\theta}_{h,\lambda}=O(h^{2})$.
%Moreover, we have \comment{could be wrong...}
%\[
%\theta_{h,\lambda}=\frac{1}{N}\sum_{i=1}^{N}\theta_{i}+O(h^{2}).
%\]
\end{prop}
\subsection{Extension of mean shape under phase variation} \label{sec:FrenetMean-warping}

We have characterized the variation of the curves $x_i(t) = X_i(s_i(t))$ in terms of its geometry using curvature and torsion, under $M_1$ in (\ref{eq:xModel}). On the other hand, the shape variation of the curves is often viewed as a curve registration problem \citep[e.g.,][]{Marron2015, Carroll2020}. 
For curves $x_0$ and $x_1$, the registration problem is motivated by finding the most appropriate warping $h:[0,T] \rightarrow [0, T]$ such that two curves $x_1(h(t))$ and $x_0(t)$ looks similar. 
In order to incorporate the registration in our framework with arclength parametrized curves, as $M_2$ in (\ref{eq:xModel}), we consider
\[
x_0(t) = X_0(s_0(t)) \,,\quad x_1(h_1(t)) = X_1(s_1(h_1(t))) \,.
\]
The corresponding Frenet paths are $Q_0(s_0(t))$ and $Q_1(s_1(h_1(t)))$. Under our framework, it is natural to express the registration problem in terms of the Frenet paths. 

Define a space warping diffeomorphism $\gamma:[0, L_0] \rightarrow [0, L_1]$ for any $h\in H_T$ such that  $s_1\circ h = \gamma \circ s_0$. Denote the function space of the space warping diffeomorphisms by $\Gamma_S$. Then, the warping problem in $x$ is translated into that of $Q_0(s)$ and $Q_1(\gamma(s))$. That is, for two Frenet paths $Q_{0}\,:\,\left[0,L_{0}\right]\longrightarrow SO(p)$
and $Q_{1}\,:\,\left[0,L_{1}\right]\longrightarrow SO(p)$, the curves are stretched using a diffeomorphism $\gamma\,:\,\left[0,L_{0}\right]\longrightarrow\left[0,L_{1}\right]$. For length normalised curves, we have $L_0=L_1=1$, see section \ref{sec:scaling}.
 
From section \ref{sec:Frenet}, the Frenet path $s\mapsto\tilde{Q}_{1}(s)=Q_{1}(\gamma(s))$ 
is also the solution of the following Frenet-Serret ODE:
\begin{eqnarray*}
\frac{d}{ds}\tilde{Q}_{1}(s) & = & Q_{1}'(\gamma(s))\gamma'(s) 
  =  Q_{1}\left(\gamma(s)\right)A_{\theta}(\gamma(s))\gamma'(s)
 =  \tilde{Q}_{1}(s)A_{\tilde{\theta}}(s) \,,
\end{eqnarray*}
where 
\begin{equation} \label{eq:theta-tilde}
    \tilde{\theta}(s)=\theta(\gamma(s))\gamma'(s) \,.
\end{equation} 
%The curve $X_{1}$ is stretched nonlinearly in order to look like the geometric curve similar to $X_{0}$, and $\tilde{\theta}$ is the corresponding stretched generalised curvature.
It also satisfies the self-prediction property: for all $t, s \in [0,1]$ such that $|t-s|\leq 1$, we have
\begin{align*}
\tilde Q_1(t) &= \phi_{\tilde\theta}(t-s, s, \tilde Q_1(s)) \,.
%&= \phi_\theta(\gamma(t)-\gamma(s), \gamma(s), Q(\gamma(s)) \,.
\end{align*}
The self-prediction error criterion (\ref{eq:selfpredCrt}) for $\tilde Q_i$ can be expressed as
\begin{align*}
\mathcal{V}(\tilde Q_1, \phi_{\tilde\theta_1}) &=\int_0^1\int_0^1 d\left(Q_1(\gamma(t)), \phi_{\theta(\gamma)\gamma^\prime}(t, s, Q_1(\gamma(s)))\right)^2\,dsdt \\
&= \int_0^1\int_0^1 d\left(Q_1(\gamma(t)), \phi_\theta(\gamma(t)-\gamma(s), \gamma(s), Q_1(\gamma(s))\right)^2\,dsdt \,. %\\
%&= \int_0^1\int_0^1 d\left(Q(t)), \phi_\theta(t - s, s, Q(s))) (\gamma^{-1})^\prime(s)(\gamma^{-1})^\prime(t)\right)^2\,dsdt \,.
\end{align*}
We define an augmented self-prediction error criterion by
\[
\mathcal{V}_2(Q, \phi_\theta) = \inf_{\gamma\in \Gamma_S}\mathcal{V}(Q(\gamma), \phi_{\theta(\gamma)\gamma^\prime}) \,.
\]
\begin{definition}\label{def:MeanShape2}
Assume that $Q_i, i=1,\ldots, N$ be the independent and identically distributed random Frenet paths with the same distribution as $Q$, associated with parameters $\theta_i\in\mathcal{H}$, satisfying $Q_i^\prime = Q_i A_{\theta_i}$. For random space diffeomorphisms $\gamma_i \in \Gamma_S$ identically distributed as $\gamma$, let $\tilde Q_i = Q_i(\gamma_i)$ be the observed Frenet paths. The mean parameter for $\tilde{Q}$ is defined as
\[
\bar\theta_2  = \arg\min_{\theta\in \mathcal{H}} E\{ \inf_{\gamma\in \Gamma_S}\mathcal{V}(Q(\gamma), \phi_{\theta(\gamma)\gamma^\prime})\} = \arg\min_{\theta\in\mathcal{H}}E\{\mathcal{V}_2(Q, \phi_\theta)\} \,.
\]
\end{definition}
The relation (\ref{eq:theta-tilde}) defines a ``spatial'' or geometric registration based on the family of deformations defined as $\theta\mapsto \gamma\cdot\theta = \gamma^{\prime}\theta\circ \gamma$,
for any increasing diffeomorphism $\gamma$ . This is a group action, i.e. for all $\gamma_{1},\gamma_{2}$ diffeormophisms, and any generalised curvature $\theta$, we have
\[
\left(\gamma_{2}\circ \gamma_{1}\right)\cdot\theta=\gamma_{2}\cdot\left(\gamma_{1}\cdot\theta\right) \,.
\]
Note that the stretching action by warping does not permit to transform any geometry into another. Indeed, if $\theta_{0}$ and $\theta_{1}$ are two generalised curvatures such that the torsion $\tau_{0}>0$ and $\tau_{1}<0$, then we cannot find $\gamma$ such that $\gamma^{\prime}\tau_{1}(\gamma)=\tau_{0}$ \citep{BrunelPark2019}. Our mean parameter is identified as the solution to a constrained minimization problem.  

%The derived relation (\ref{eq:theta-tilde}) can be contrasted with the usual assumption for warping derivative curves by $\tilde\theta = \theta(\gamma)$ \citep[e.g.,][]{RamsayLi1998, Sangalli2009b}. In our experiences, the former fits better the empirical data than the latter.

%%
\section{Estimation algorithm} \label{sec:estimation}

%We have seen that the main estimation problem is formulated in terms of the Frenet paths $Q$. 
We first derive a main algorithm for estimation under shape variation model and extend it to cover phase variation model. As the Frenet paths are often not directly available, we suggest pre-processing methods to treat the Euclidean curves. Furthermore, as a special case of the Euclidean curves with a manifold structure, we show that our estimation algorithm can be applied to the spherical curves by reparametrization.

\subsection{Estimation under shape variation} \label{sec:estShape}

Based on the statistical criterion developed in section~\ref{sec:Stat-crt}, the estimation of $\mbox{\ensuremath{\theta}}$ from given Frenet paths
$\boldsymbol{Q}$ is done by solving
\begin{equation}\label{eq:crt_theta}
\widehat{\theta}_{h,\lambda}\left(\cdot\right)=\arg\min_{\theta}\mathcal{I}_{h,\lambda}\left(\theta\right) \,.
\end{equation}
We need to solve the nonparametric estimation problem (\ref{eq:crt_theta}),
but in practice, we solve this by discretization. 
Suppose that $Q_i, i=1,\ldots, N$ is available at finite grid points $s_{ij}, j=1,\ldots, n_i$. 
%We write $U_{ij}=Q_i(s_{ij})$ for the observations. 
Then we discretize the integral on a grid $0=t_{i1}<t_{i2}<\dots<t_{iQ_{i}}=1$
and minimize 
{\footnotesize\[
\min_{\theta\in\mathcal{H}}\sum_{i=1}^{N}\frac{1}{n_{i}Q_i}\sum_{j,q=1}^{n_{i},Q_{i}}K_{h}(t_{iq}-s_{ij})\left\Vert \log\left(Q_{i}(t_{iq}){}^{\top}Q_{i}(s_{ij})\exp\left(\left(t_{iq}-s_{ij}\right)A_{\theta}\left(\frac{t_{iq}+s_{ij}}{2}\right)\right)\right)\right\Vert _{F}^{2}+\lambda\| \theta^{''}\|_2^{2}
\]}
The presence of the exponential makes the optimization difficult,
and we use an additional approximation that provides a simple algorithm and simplifies the analysis of our estimator. 
Let 
\[
L_{ijq} = \log\left(Q_{i}(t_{iq}){}^{\top}Q_{i}(s_{ij})\exp\left(\left(t_{iq}-s_{ij}\right)A_{\theta}\left(\frac{t_{iq}+s_{ij}}{2}\right)\right)\right) \,.
\]
Define $u_{ijq}=t_{iq}-s_{ij}$, $v_{ijq}=\frac{t_{iq}+s_{ij}}{2}$. The first term in the criterion can be expressed as
\[
\sum_{i,j,q=1}^{N,n_{i},Q_{i}}\frac{1}{n_{i}Q_{i}}K_{h}(u_{ijq})\Vert L_{ijq}\Vert_F^2 \,.
\]
We define the skew-symmetric matrix ${R}_{ijq}=-\frac{1}{t_{iq}-s_{ij}}\log\left(Q_{i}(t_{iq})^{\top}Q_{i}(s_{ij})\right)$. 
We derive a first order approximation to $\Vert L_{ijq}\Vert_F$ based on the Baker-Campbell-Hausdorff formula \citep{Higham2008}: for $t$ small enough,
\begin{equation*}
\exp(tA)\exp(tB)=\exp\left(tA+tB+\frac{1}{2}t^{2}\left[A,B\right]+O(t^{3})\right) \,.\label{eq:BCHformula}
\end{equation*}
where $[A, B] = AB - BA$. 
In particular, it can be shown (S2.2 in the supplementary) that
\[
%& \sum_{i=1}^{N}\frac{1}{n_{i}}\sum_{j,q=1}^{n_{i},Q_{i}}K_{h}(t_{iq}-s_{ij})\left\Vert \log\left(Q_{i}(t_{iq}){}^{\top}Q_{i}(s_{ij})\exp\left(\left(t_{iq}-s_{ij}\right)A_{\theta}\left(\frac{t_{iq}+s_{ij}}{2}\right)\right)\right)\right\Vert _{F}^{2} \\ 
\sum_{i,j,q=1}^{N,n_{i},Q_{i}}\frac{1}{n_{i}Q_{i}}K_{h}(u_{ijq})\Vert L_{ijq}\Vert_F^2 = \sum_{i,j,q=1}^{N,n_{i},Q_{i}}\frac{1}{n_{i}Q_{i}}K_{h}(u_{ijq})u_{ijq}^{2}\| A_{\theta}(v_{ijq})-{R}_{ijq}\| _{F}^{2} + O(h^3) \,.
\]
This motivates us to introduce a new approximate criterion, %with $u_{ijq}=t_{iq}-s_{ij}$, $v_{ijq}=\frac{t_{iq}+s_{ij}}{2}$
\[
\tilde{\mathcal{I}}_{h,\lambda}\left(\theta; \bm{R}\right)=\sum_{i,j,q=1}^{N,n_{i},Q_{i}}\frac{1}{n_{i}Q_{i}}K_{h}(u_{ijq})u_{ijq}^{2}\| A_{\theta}(v_{ijq})-{R}_{ijq}\| _{F}^{2}+\lambda\int_{0}^{1}\| \theta^{\prime\prime}(t)\|^{2}dt \,.
\]
In the particular case of $p=3$, if we define
\[
{R}_{ijq}=\left[\begin{array}{ccc}
0 & -r_{ijq}^{1} & -r_{ijq}^{3}\\
r_{ijq}^{1} & 0 & -r_{ijq}^{2}\\
r_{ijq}^{3} & r_{ijq}^{2} & 0
\end{array}\right]
\]
the Frobenius norm can be rearranged with weights $\omega_{ijq}=\frac{2}{n_{i}Q_{i}}K_{h}(u_{ijq})u_{ijq}^{2}$
%{\footnotesize\begin{equation}
using the elementwise expansion, leading to
\[
\tilde{\mathcal{I}}_{h,\lambda}\left(\theta;  \boldsymbol{R}\right)=\sum_{i,j,q=1}^{N,n_{i},Q_{i}}\omega_{ijq}\left(\kappa(v_{ijq})-r_{ijq}^{1}\right)^{2}+\sum_{i,j,q=1}^{N,n_{i},Q_{i}}\omega_{ijq}\left(\tau(v_{ijq})-r_{ijq}^{2}\right)^{2}+\lambda\int_{0}^{1}\| \theta^{\prime\prime}(t)\|^{2}dt \,. %\label{eq:I1_g}
\]
%\end{equation}}
%\[
%\tilde{\theta} = \arg\min_{\theta \in \mathcal{H}} \sum_{i,j,q=1}^{N,n_i,Q_i} \omega_{ijq} \left ( \theta(v_{ijq}) - \tilde{r}_{ijq}^{k}) \right )^2  + \lambda \int_0^1 ||\theta''||^2 dt 
%\]
%and $\widehat{\theta}=\left(\tilde{\theta}_{1},\dots,\tilde{\theta}_{p-1}\right)^{\top}$.
That is, the optimization problem for $p=3$
\[
\tilde{\theta}_{h,\lambda}=\arg\min_{\theta\in\mathcal{H}}\tilde{\mathcal{I}}_{h,\lambda}\left(\theta; \boldsymbol{R}\right)
\]
gives rise to the computation of $2$ independent smoothing splines
(with splines of third order), defined at the knots $v_{ijq}$, with
the pseudo-observations $r_{ijq}^{1},r_{ijq}^{2}$. %,\dots r_{ijq}^{p-1}$.
The only difference with respect to the classical smoothing splines
is the presence of the weights $\omega_{ijq}$. 

\begin{rem}
Our prediction error depends on $h,\lambda=(\lambda_1,\lambda_2)$. %and we can do cross-validation.
If $h$ is too big, we integrate along the whole interval and the errors
accumulate, and it is better to restrict to smaller interval. We consider
the prediction of a small percentage (10\%, $h\approx0.1$) of the
individuals, when the total length of a curve is 1. 
 In our numerical studies we have performed 10-fold cross validation by minimizing
\[
%\widehat{\mathcal{J}_{K}}(h,\lambda_{1},\lambda_{2})=
\sum_{k=1}^{K}\sum_{(i,j)\in T_{k}}\left\|\log\left(U_{ij}^\top \hat Q_i^{-(k)}(s_{ij}; h,\lambda)\right)\right\|_F^2
%d\left(\hat{M}_i^{(-k)}(s_{ij}),U_{ij}^{(k)}\right)
\]
where $T_k$ is the $k$th index set based on $K=10$ random partition of the observations $\left\{U_{ij} = Q_i(s_{ij})\right\}$,  $i=1,\ldots,N, j=1, \ldots,n_i$ and $\hat{Q}_{i}^{-(k)}(s_{ij}; h,\lambda)$ are the predicted Frenet paths reconstructed with parameters $\hat\theta^{-(k)}$, estimated without the $k$th partition dataset, using hyperparameters $h, \lambda$ and the initial value of $\bar Q(0)$.
\end{rem}

\subsection{Estimation under phase variation} \label{sec:est-phase}

Under the phase variation model $M_2$ in (\ref{eq:xModel}), the mean parameter needs to be refined to satisfy (\ref{eq:theta-tilde}). This is translated into the problem of aligning the raw estimates of the parameters ${r}_{ijq} = ({r}_{ijq}^1, {r}_{ijq}^2)$ to obtain the optimal warping function $\gamma_i$. 
%\comment{Better to use the initial mean estimate and iterate?}
Define $\tilde{R}_{ijq} = \gamma_i(v_{ijq})\cdot R_{ijq}$ as in (\ref{eq:theta-tilde}). Then, the estimates are defined as the minimizer of
\[
%\tilde{\theta}_{h,\lambda} = \arg\min_{\theta \in \mathcal{H}}
\tilde{\mathcal{I}}_{h,\lambda}(\theta; \tilde{\boldsymbol{R}}) =
\sum_{i,j,q=1}^{N,n_{i},Q_{i}}\omega_{ijq}\left(\kappa(v_{ijq})-\tilde{r}_{ijq}^{1}\right)^{2}+\sum_{i,j,q=1}^{N,n_{i},Q_{i}}\omega_{ijq}\left(\tau(v_{ijq})-\tilde{r}_{ijq}^{2}\right)^{2}+\lambda\int_{0}^{1}\| \theta^{\prime\prime}(t)\|^{2}dt \,. %\label{eq:I1_g}
\]
%and $\widehat{\theta}=\left(\tilde{\theta}_{1},\dots,\tilde{\theta}_{p-1}\right)^{\top}$.
For the alignment of the raw estimates, we implement a version of the iterative algorithm similar to those developed in \citet{KneipRamsay08, Tucker2013} based on the Kahrunen-Lo\`{e}ve expansion : $r_i\circ \gamma_i \approx \nu + \sum_{k=1}^K\xi_{ik}\phi_k$ where $\phi_k$ are the functional principal components and $\xi_{ik}$ are the corresponding scores. The alignment algorithm is summarized below. The main difference is in step 2 to satisfy (\ref{eq:theta-tilde}) with multiplication factor $\dot{\gamma}$ instead of $\sqrt{\dot{\gamma}}$ as in (\ref{eq:DistanceSRVF}).

\paragraph{Alignment algorithm\label{par:Estimation-Algorithm3}}
Given observations $(r_i)_{i=1,\ldots,N}$, set the initial values $y_i^0 = r_i, \gamma^0 = id$, $\nu^0 = \sum_i\omega_i y_i^0$.
For $\ell\geq 1$, the optimal warping functions $(\gamma_i)_{i=1,\ldots,N}$ are found by iterating the following steps until convergence: 
\begin{enumerate}
\item Refine $y_i$: $\tilde y^{(\ell)} = \nu^{(\ell-1)} + PCAAPPROX(y^{(\ell-1)}-\nu^{(\ell-1)}, K)$
\item Update $\gamma$: $\gamma_i^{(\ell)} = \arg\min_\gamma \| \tilde y_i^{(\ell)} - (y_i^{(\ell-1)}\circ \gamma)\dot{\gamma}\|_{2}$ for $i=1,\ldots, N$
\item Update $y$: $y_i^{(\ell)} = (y_i^{(\ell-1)}\circ \gamma_i^{(\ell)})\dot{\gamma}_i^{(\ell)}$ for $i=1,\ldots, N$
\item Update $\nu$: $\nu^{(\ell)} = \sum_{i} \omega_i y_i^{(\ell)}$ 
\end{enumerate}

\subsection{Estimation from noisy Euclidean curves} \label{sec:preprocessingX}
%%
%\comment{combine with the numerical study ?}

The Frenet paths $Q$ are usually derived from Euclidean curves by pre-processing.
Suppose that the noisy observations $y_j \in \mathbb{R}^3$ satisfy $y_{j}=X(s_{j})+\sigma\epsilon_{j}, j=1,\dots,n$,
%\begin{equation}
%y_{j}=X(s_{j})+\sigma\epsilon_{j},\:j=1,\dots,n\,. \label{eq:DGP_X_SIngle}
%\end{equation}
where $X(s)$ has a Frenet path $Q(s)$ solution of the ODE $Q^{\prime}(s)=A_{\theta}(s)Q(s)$.
%We consider sample sizes $n=100,\,200$ and $\sigma=0.05$ or $\sigma=0.02$. 
We assume that the arclength parametrization can be done relatively easily, by a simple estimate of the first derivative. An added difficulty with this setting is related to defining a preliminary estimate of the Frenet path. 
As a preprocessing step, we nonparametrically estimate the higher-order derivatives of $X$, $X^{(k)}, k=1,2,3$ from the noisy observations $y_{1},\dots,y_{n}$. These derivatives can be very noisy and are used for computing raw estimates $U_{j} = \hat{Q}(s_{j}),\,j=1,\dots,n$. We consider two methods for deriving these estimates: 
\begin{description}
\item [{$\hat{Q}^{GS}(s)$}] obtained by Gram-Schmidt orthonormalization
of the frame $\left[X^{(1)}\vert X^{(2)}\vert X^{(3)}\right]$. The
derivatives are estimated by a standard local polynomial of order
4. With the same derivative estimates, we can compute the estimators
of the curvature $\hat{\kappa}^{Ext}$ and torsion $\hat{\tau}^{Ext}$
using the extrinsic formulas.
\item [{$\hat{Q}^{LP}(s)$}] obtained by constrained
nonparametric smoothing of $X$. Instead of the standard local polynomial,
we use a local expansion that uses the orthogonal vectors $T,N,B$:
{\small\[
X(s+h)=X(s)+\left(h-\frac{h^{3}\kappa^{2}(s)}{6}\right)T(s)+\left(\frac{h^{2}\kappa(0)}{2}+\frac{h^{3}\kappa^{\prime}(s)}{6}\right)N(s)+\frac{h^{3}\kappa(s)\tau(s)}{6}B(s)+o(s^{3})
\]}
\end{description}
We find in our numerical studies that $\hat Q^{LP}$ outperforms so this is used to construct our estimator.
%$\hat{Q}^{FS}$ as well as $\hat{\kappa}^{FS}$ and $\hat{\tau}^{FS}$. 
%The estimation error is accounted in our statistical model (\ref{eq:StatisticalModelNoise}) and these initial estimates are fed into our estimation algorithm for $U$. 

%%
\subsection{Estimation of curves on the sphere} \label{sec:sphericalcurve}

Our formulation does not require specific structure on the Euclidean curves. Nevertheless, it is of interest if our method is applicable to a structured data such as curves on a manifold. Of course, it is possible to estimate the curvature and torsion without additional knowledge on the manifold. However, since curvature and torsion for spherical curves are intrinsically related, direct estimation does not necessarily respect the constraints, but a constrained optimization is not obvious in this setting either.
It turns out that, instead of modifying the algorithm, we can reformulate the problem under our Frenet framework for the spherical curves.
%, demonstrating a possible extension of our method to the case of curves on manifolds. 
%To illustrate the possible extension of our method to the case of curves on manifolds we derive a reformulation of the problem under our Frenet framework for the special case of curves on the sphere. 
%We derive a Frenet-Serret formula for spherical curves.

We consider a curve $\alpha$ on a sphere of radius $R$ and center $(0,0,0)$. By definition we have $\|\alpha(t)\| = R $ for all $t \in [0, T]$. 
We consider now the curve parametrised by arc length. 
As for all $s \in [0, L], \|\alpha(s)\| = R$, we have $\langle\alpha(s), \alpha(s)\rangle = R^2$ so $2\langle\alpha(s), \alpha'(s)\rangle = 0$, thus $\alpha(s)$ is orthogonal to $\alpha'(s)$ for all $s$. We denote $\beta := \alpha'$. 
We define the spherical unit normal as $\gamma(s) = (1/{R}) \alpha(s) \wedge \beta(s).$
Since $\|\alpha(s)\|/{R} = 1 = \|\beta(s)\|$ for all $s$ and the two are orthogonal, $\|\gamma(s)\|=1$ too.

\begin{definition}
Define the \textit{geodesic curvature} of a spherical curve $\alpha: [0, T] \longrightarrow \mathbb{S}^2$ parametrised by arclength to be 
$$k_g(s) = \langle\alpha''(s),  \gamma(s)\rangle \,.$$
The \textit{geodesic curvature} measures the failure of a curve to be a geodesic. 
\end{definition}

\begin{prop}{(Frenet-Serret formula for spherical frames)}\label{prop:FrenetS2} Let $\alpha: [0, T] \longrightarrow \mathbb{S}^2$, unit sphere, be a spherical curve parametrised by arclength. Let $\beta(s) = \alpha'(s)$ and $\gamma(s) = \alpha(s) \wedge \beta(s)$. The vectors ({$\alpha, \beta, \gamma$}) define the spherical frame and satisfy the following equation with $k_g(s)= \langle\alpha''(s), \gamma(s)\rangle$
% We define the geodesic curvature as $k_g(s)= \langle\alpha''(s), \gamma(s)\rangle$. The derivative of the frame at $s \in [0, 1]$ is given in terms of the geodesic curvature $k_g(s)$ by:
\[ 
\begin{cases}
\alpha'(s) = \beta(s) \\
\beta'(s) = -\alpha(s) + k_g(s)\gamma(s) \\
\gamma'(s) = -k_g(s) \beta(s)
\end{cases}
\]
%where $k_g(s)= \langle\alpha''(s), \gamma(s)\rangle$ is the geodesic curvature.
\end{prop}
This proposition implies that if one knows the initial position and direction, a given geodesic curvature function $k_g(s)$ determines a unique spherical curve parametrised by arclength. Therefore, we can directly apply our algorithm with the Frenet frame for spherical curves to obtain an estimate of the geodesic curvature $\hat{k}_g$. Then we reconstruct the curve by solving the spherical Frenet-Serret ODE above. This method ensures that the estimated mean is in $\mathbb{S}^2$.
%A numerical example is included in section~\ref{sec:numeric}.
%We obtained very satisfactory results by applying this framework to curves on the sphere generated by the model detailed in \cite{DaiMuller}. \comment{? maybe add some results or remove this sentence}

%Although the reformulation is specific to the case of curves on a sphere, it gives an indication of the potential for extension of our method and we envisage that a similar representation could be derived for general manifolds under the so-called Darboux frame.  

%%%%%%%%%%%%%%%%%%%%%%% 
\section{Numerical studies} 
\label{sec:numeric}

We conduct simulation studies to assess performance of the proposed methods in identifying mean geometry (curvature, torsion) as well as mean shape in finite samples, followed by real data examples. 
%The main components to evaluate are the quality of the mean parameter estimation under noise, the effect of the pre-processing in the case of indirectly observed Frenet paths and the mean shape estimation. 
%Where possible, we compare our results to the elastic shape analysis based on the SRVF framework described in \ref{sec:elastic}, using python package \texttt{fdasrsf} \citep{fdasrsf} and the standard functional arithmetic mean. 
%For the former, we use a \texttt{Python} implementation available in the package \texttt{fdasrsf} \citep{fdasrsf}. 

%Numerical summaries of $L^2$ distance as well as Fisher-Rao distance (\ref{eq:DistanceSRVF}) are compared (Table B. in the supplementary). 

%%
\subsection{Data generating process}
%We consider the estimation problem with directly observed Frenet paths and indirectly observed Frenet paths from Euclidean curves under a reference model. 

We consider the cases of direct observations of Frenet paths and indirect observations from Euclidean curves, possibly contaminated by noise.
These are studied under both shape variation (S1) and phase variation (S2) models as defined in (\ref{eq:xModel}). Since the mean parameter is not always available, we add a case of Euclidean curves with unknown parameters (S3). We also include an example of spherical curves as a special of Euclidean curves with a manifold structure (S4).

\subsubsection{Scenario 1: Shape variation model}
The reference shape is defined by $\bar\kappa(s)= \exp(\zeta\sin(s)), 
\bar\tau(s)= \eta s-0.5$ with $\zeta=1,\,\eta=0.2$ for  $s \in [0, 5]$.
%\[
%\begin{cases}
%\bar\kappa(s)= & \exp(\zeta\sin(s))\\
%\bar\tau(s)= & \eta s-0.5
%\end{cases},\:\zeta=1,\,\eta=0.2 \,.
%\]
%and we set $s \in [0, 5]$. 
We simulate a population of random Frenet paths $s\mapsto Q_{i}(s),\:i=1,\dots,N$ generated by random Frenet-Serret equations with random individual shape parameter $\theta_{i}=(\kappa_i, \tau_i)$ obtained as $\kappa_{i}= \left|\bar{\kappa}+\sigma_{\kappa}\zeta_{i}^{^{1}}\right|, \tau_{i}= \bar{\tau}+\sigma_{\tau}\zeta_{i}^{2}$,
%\begin{equation*}
%\begin{cases}
%\kappa_{i}= & \left|\bar{\kappa}+\sigma_{\kappa}\zeta_{i}^{^{1}}\right|\\
%\tau_{i}= & \bar{\tau}+\sigma_{\tau}\zeta_{i}^{2}
%\end{cases}\label{eq:RandomTheta}
%\end{equation*}
where $\zeta_{i}^{^{1}},\,\zeta_{i}^{2}\;i=1,\dots,N$ are centered independent Gaussian processes with (unit) Mat\'ern covariance functions\footnote{$k(s,s')=\frac{1}{\Gamma(\nu)2^{\nu-1}}\left(\frac{\sqrt{2\nu}}{\ell}\left|s-s'\right|\right)^{\nu}K_{\nu}\left(\frac{\sqrt{2\nu}}{\ell}\left|s-s'\right|\right)$} with $\nu=\frac{5}{2}$ and characteristic length scale $\ell=1$. We set $\sigma_{\kappa}=\sigma_{\tau}=0.3$. This means that the random functions are twice differentiable, and the functions $\bar{\kappa},\bar{\tau}$ are respectively the means of the population $\left(\kappa_{i}\right)_{i=1\dots N}$ and $\left(\tau_{i}\right)_{i=1\dots N}$. 
For the Frenet paths, we allow for random initial conditions $Q_i(0) = Q_i^0$ where $Q_{i}^{0}\sim\mathcal{F}(I_{3},\alpha_0)$ with $\alpha_0=10$, Fisher-Langevin distribution with mean identity and concentration $\alpha_0=10$.
%, and its density given by
%$f(U;D,\alpha)=\left\{ _{0}F_{1}(\frac{p}{2};\frac{\alpha^{2}}{4})\right\} ^{-1}\exp\left(\alpha\mbox{Tr}\left(D^{\top}U\right)\right)$.
Denote by $X_{\theta_i}$ the corresponding Euclidean curves to $Q_i$.
% The number of observations $n_{i}$ per curve is $n_{i}=100$. 
We consider two types of observations models:
\begin{description}
\item [S1.1] Observations as Frenet paths:
\[
U_{ij} = Q_i(s_{ij})M_{ij}\,,\quad i=1,\ldots, N,\, j=1,\ldots, n_i \,,
\]
where random rotations $M_{ij}\sim\mathcal{F}(I_{3},\alpha)$. 
\item [S1.2] Observations as noisy Euclidean curves:
\[
y_{ij}=X_{\theta_i}(s_{ij})+\sigma_{e}\epsilon_{ij}\,,\quad  i=1,\ldots,N, j=1,\ldots,n_i \,.
\]
\end{description}

\subsubsection{Scenario 2: Shape and Phase variation model}
The reference shape is defined by $\bar\kappa(s)= 10(\sin(3s) + 1), \bar\tau(s)= -10\sin(2\pi s)$
%\[
%\begin{cases}
%\bar\kappa(s)= & 10(\sin(3s) + 1) \\
%\bar\tau(s)= & -10\sin(2\pi s)
%\end{cases}
%\]
and we set $L=1$ and $\bar{s}(t) = t$. We simulate a population of Frenet paths $s\mapsto Q_{i}(s),\:i=1,\dots,N$ generated by Frenet-Serret equations with individual shape parameter $\theta_{i}$ obtained as obtained as $\kappa_{i}= \omega_i'(s) \bar{\kappa}(\omega_i(s)), \tau_{i}= \omega_i'(s) \bar{\tau}(\omega_i(s)),$ 
%\begin{equation*}
%\begin{cases}
%\kappa_{i}= & \omega_i'(s) \bar{\kappa}(\omega_i(s)) \\
%\tau_{i}= & \omega_i'(s) \bar{\tau}(\omega_i(s)) 
%\end{cases}\label{eq:WarpingTheta}
%\end{equation*}
where $\omega_i(s) = \frac{\log(s (\exp(a_i)-1)+1)}{a_i}$ if $a_i \neq 0$ otherwise $\omega_i(s)=s$, and their inverse functions $\omega_i^{(-1)}(s) = \gamma_i(s) = \frac{\exp(a_i s) - 1}{\exp(a_i) - 1}$ define the space warping functions. We choose $a_i$ equally spaced between $-1$ and $1$. 
%Similarly to Scenario 1, we define two types of observation models with Frenet paths (\textbf{S2.1}) and noisy Euclidean curves (\textbf{S2.2}) with $s_{ij} = \bar{s}_j$.
Similarly to S1, we consider observations as Frenet paths (\textbf{S2.1}) and Euclidean curves (\textbf{S2.2}) with $s_{ij} = \bar s_j$.
%We consider three types of observations models. The first two are comparable to Scenario 1 and 
In addition, we add \textbf{S2.3} to emulate time warping in the Euclidean curves.
%We consider three types of observations models. The first two are comparable to Scenario 1 and the third one is added to emulate time warping in the Euclidean curves.
\begin{description}
%\item [S2.1] Observations as Frenet paths:
%\[
%U_{ij} = Q_i(\bar{s}_{j})M_{ij}\,,\quad i=1,\ldots, N,\, j=1,\ldots, n_i \,,
%\]
%where random rotations $M_{ij}\sim\mathcal{F}(I_{3},\alpha)$. 
%\item [S2.2] Observations as noisy Euclidean curves:
%\[
%y_{ij}=X_{\theta_i}(\bar{s}_{j})+\sigma_{e}\epsilon_{ij}\,,\quad  i=1,\ldots,N, j=1,\ldots,n_i \,.
%\]
%The $X$ curves are already arclength parametrized and the mean parameters $\bar{\kappa},\bar{\tau}$ define the mean shape corresponding to the population of Euclidean curves $X_{i}=X_{\theta_i},\:i=1,\dots,N$.
\item [S2.3] Observations as noisy Euclidean Curves with additional time warping:
%Inspired by \citet{BrunelPark2019}, 
We generate the arclength functions according to
\[
s_i(t) = \gamma_i \circ \bar{s} \circ h_i (t)\,,\quad i=1,\ldots, N \,,
\]
where $h_i(t) = b_i \sin(2\pi t) + t$ and $b_i$ are equally spaced between $-0.1$ and $0.1$ so that the functions remain strictly increasing. %and define warping functions. 
We have $X_{\theta_i}(s_{ij}) = x_{\theta_i}(t_j)$ with $t_0 < ... < t_n$ equally spaced between $0$ and $1$ and the measurement model is defined as
\[
y_{ij}=x_{\theta_i}(t_{j})+\sigma_{e}\epsilon_{ij}\, \quad i=1,\ldots,N, j=1,\ldots,n_i \,.
\]
\end{description}
In S1 and S2, $\alpha$ and $\sigma_e$ control the noise level in the data respectively.

\subsubsection{Scenario 3: Model with unknown parameters} \label{sec:simS3}

%In this section, we investigate  the  case  when  the  true  mean  parameter is  implicitly  defined.  
We treat  the  case  where  the  true  mean  parameter is  implicitly defined. 
We consider a parametric curve defined by $x_{1}(t)=\cos(at),$ $x_{2}(t)=\sin(bt),$ and $x_{3}(t)=ct$, for $t\in\left[0,5\right]$. We denote by $\varphi=(a,b,c)$ the parameter. The corresponding curvature and torsion are parametric functions of $\varphi$. Individual parameters are simulated from ${\varphi}_{i}\sim\mathcal{N}\left({\varphi}_{ref},\sigma_{P}\right)$ where $\sigma_{P}>0$ is the population variability and ${\varphi}_{ref}=(1,.9,.8)$. 
%, see Figure \ref{fig:exInitUnknownParam} for the individual curves under $\sigma_P=0.05$.
The corresponding curvature and torsion are denoted by $\theta_{ref}$ and $\theta_i, i = 1, \ldots, N$, respectively. 
We define the population parameter as $\bar{\theta}\triangleq\frac{1}{N}\sum_{i=1}^{N}\theta_{i}$ on $\left[0,1\right]$, and because of the nonlinearity, we have $\bar{\theta}\neq\theta_{ref}$ in general. 
%compared in Figure~\ref{fig:exInitUnknownParam} \comment{move to supplemennatry}. 
Nevertheless, when $\sigma_{P}$ is relatively small (i.e lower than $0.05$ in our case), the geometry of the curves varies but the main features are preserved, meaning the curvatures $\theta_{i}$ varies around $\theta_{ref}$, such that $\theta_{ref}\approx\bar{\theta}$.
The measurements are then obtained from $y_{ij}=x_{i}(t_{j})+\sigma_{e}\epsilon_{ij},  i=1,\ldots, N, j=1,\ldots,n_i$.
%\[
%y_{ij}=x_{i}(t_{j})+\sigma_{e}\epsilon_{ij} \,,\quad i=1,\ldots, N \,, j=1,\ldots,n_i\,.
%\]
In the simulation, we vary the model by $\sigma_{P}^{2}=0.02$ (\textbf{S3.1}) or $\sigma_{P}^{2}=0.05$ (\textbf{S3.2}) and the noise level by $\sigma_{e}^{2}=0$ or $\sigma_{e}^{2}=0.03$. An example of curves is shown in the supplementary (Figure A).

%We compare the mean parameter estimates $\hat{\theta}^{pop}$ and $\hat{\theta}^{ind} = (1/N) \sum_{i=1}^{N}\hat\theta_i$ with $\bar\theta$. For comparison, we include the empirical mean estimate $\hat{\theta}^{ind}_{Ext}$ obtained from the individual extrinsic estimates $\hat{\theta}_{i}^{Ext}$ based on (\ref{eq:ExtrinsicCurvatures}). %computed with the extrinsic formula. 
%The hyperparameters are selected from $h\in\left(0.03,0.1\right)$ and $\lambda_{1},\lambda_{2}\in\left(1e^{-9},1e^{-3}\right)$. 

\subsubsection{Scenario 4: Model with curves on the sphere}

This scenario studies the special case of a population of curves lying on the manifold $\mathbb{S}^2$. 
We consider the generative model for curves on $\mathbb{S}^2$ described in \cite{DaiMuller2018}.
For $i=1,...,N$, the sample curves $x_i$ are generated as $x_i : [0,1] \longrightarrow \mathbb{S}^2, x_i(t) = \exp_{\mu(t)}{( \sum_{k=1}^{20} \xi_k \phi_k(t) )}$ with $\mu(t) = \exp_{[0,0,1]}{(\cos(\theta(t))\varphi(t), \sin(\theta(t))\varphi(t), 0)}$ the mean function in $\mathbb{S}^2$ and the arbitrary chosen functions $\theta(t) =  4t + \frac{1}{2}, \varphi(t) = 5(t+1)$.
%\[
%\begin{cases}
%\theta(t) =  4t + \frac{1}{2}\\
%\varphi(t) = 5(t+1) \,.
%\end{cases}
%\]
For $k=1,...,20, \xi_k$ are generated by independent Gaussian distributions with mean zero and variance $0.07^{k/2}$. The functions $\phi_k(t)$ are defined on $[0,1]$ as $\phi_k(t) = 2^{-1/2} R_t [\Phi_k(t/2), \Phi_k((t+1)/2), 0]^T$, where $R_t$ is the rotation matrix from $[0,0,1]$ to $\mu(t)$, and ${\{\Phi_k\}}_{k=1}^{20}$ is the orthonormal Legendre polynomial basis on $[0,1]$. 
The measurements are then obtained from $y_{ij}=x_{i}(t_{j})+\sigma_{e}\epsilon_{ij}, i=1,\ldots, N, j=1,\ldots,n_i$.
%\[
%y_{ij}=x_{i}(t_{j})+\sigma_{e}\epsilon_{ij} \,,\quad i=1,\ldots, N, j=1,\ldots,n_i\,.
%\]
%In the simulation, we vary the noise level by $\sigma_{e}^{2}=0$ or $\sigma_{e}^{2}=0.02$. 
For comparison, we identify the true mean parameter $\bar{k}_g = k^{\mu}_g$, as defined in section~\ref{sec:sphericalcurve} 

%In this case, only the hyperparameters $h$ and $\lambda_2$ are useful as we only need to estimate $k_g$. They were selected from $h\in\left(0.02,0.1\right)$ and $\lambda_{2}\in\left(1e^{-9},1e^{-3}\right)$. 

\medskip 
All simulation models are evaluated on a population of $N=25$ curves with $n=100$ sample points and are repeated for 100 times. We have run a Bayesian optimization algorithm \citep[e.g.,][]{Martinez-Cantin2015} with a standard 10 fold cross validation to search for the best hyperparameters $h$ and $\lambda$.

\subsection{Simulation results}

%% without using subfigure %%%%
\begin{figure}[tbhp]
\centering
\begin{tabular}{cc}
       \includegraphics[width=0.4\textwidth]{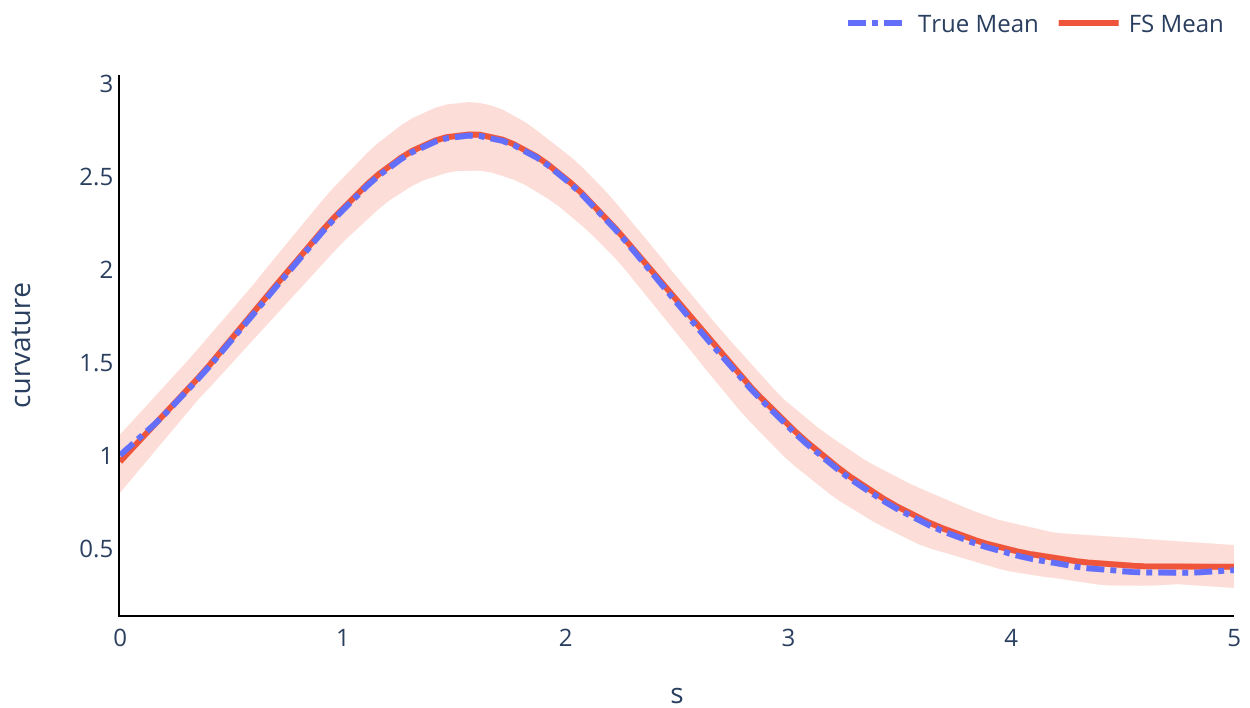} &
        \includegraphics[width=0.4\textwidth]{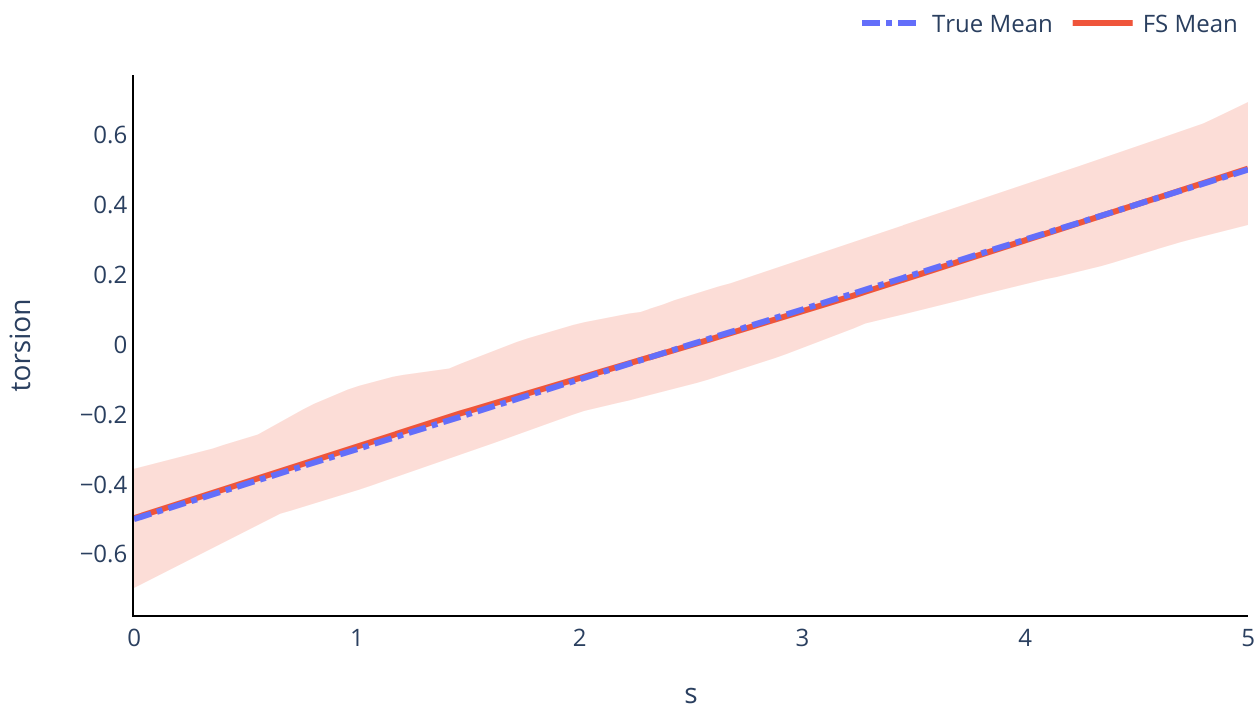} \\
        \multicolumn{2}{c}{\textbf{S1.1}} \\
        %\label{fig:Fig_S1.1}
        \includegraphics[width=0.4\textwidth]{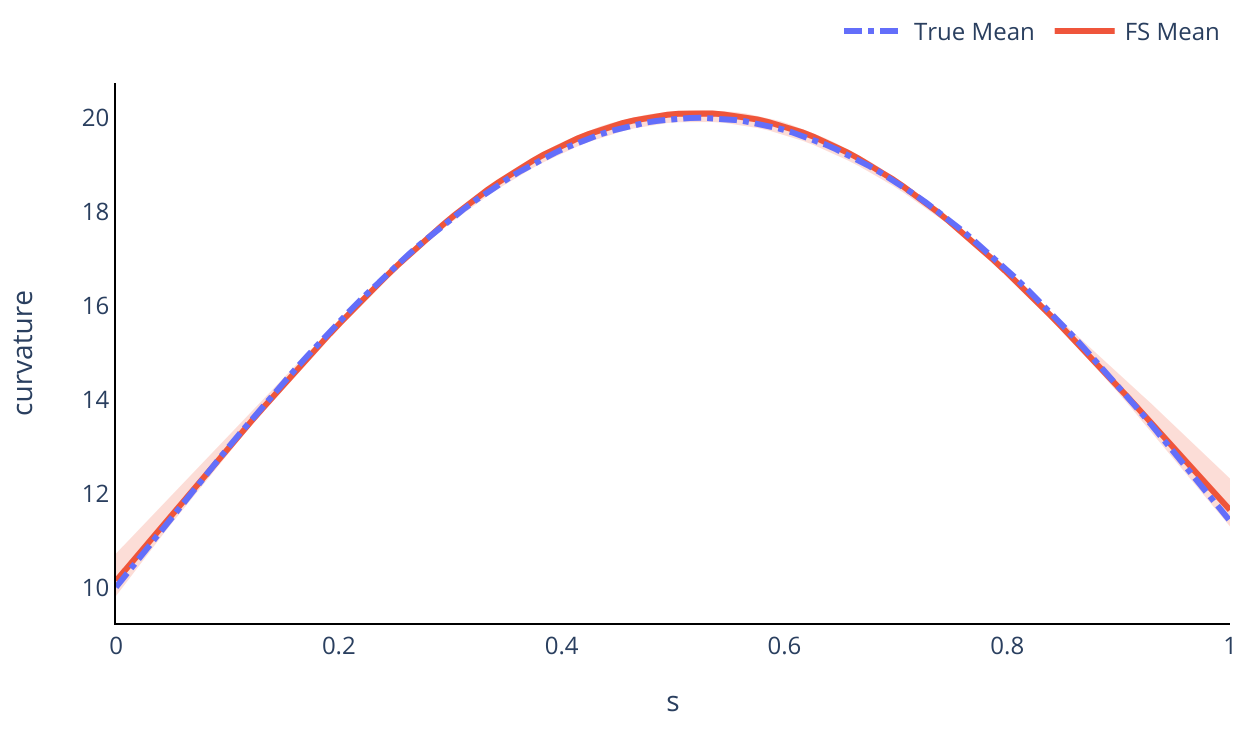} &
        \includegraphics[width=0.4\textwidth]{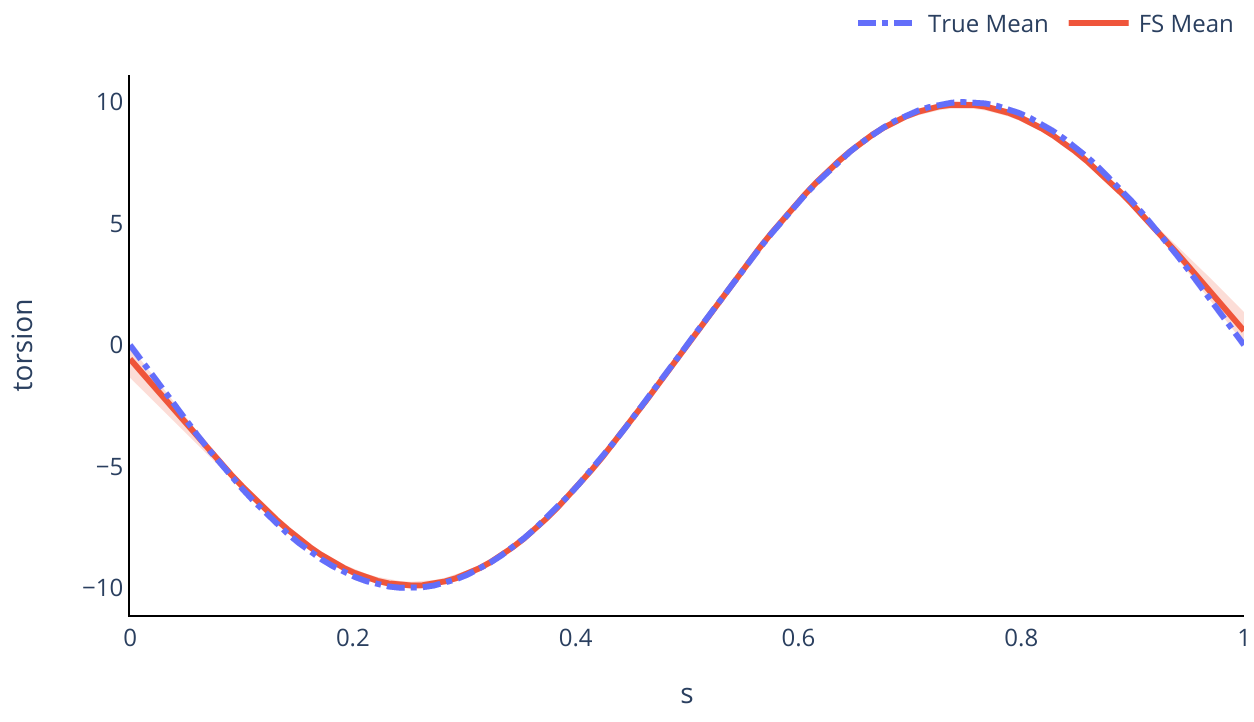} \\
        \multicolumn{2}{c}{\textbf{S2.1}}  \\
        %\label{fig:Fig_S2.1}
        \includegraphics[width=0.4\textwidth]{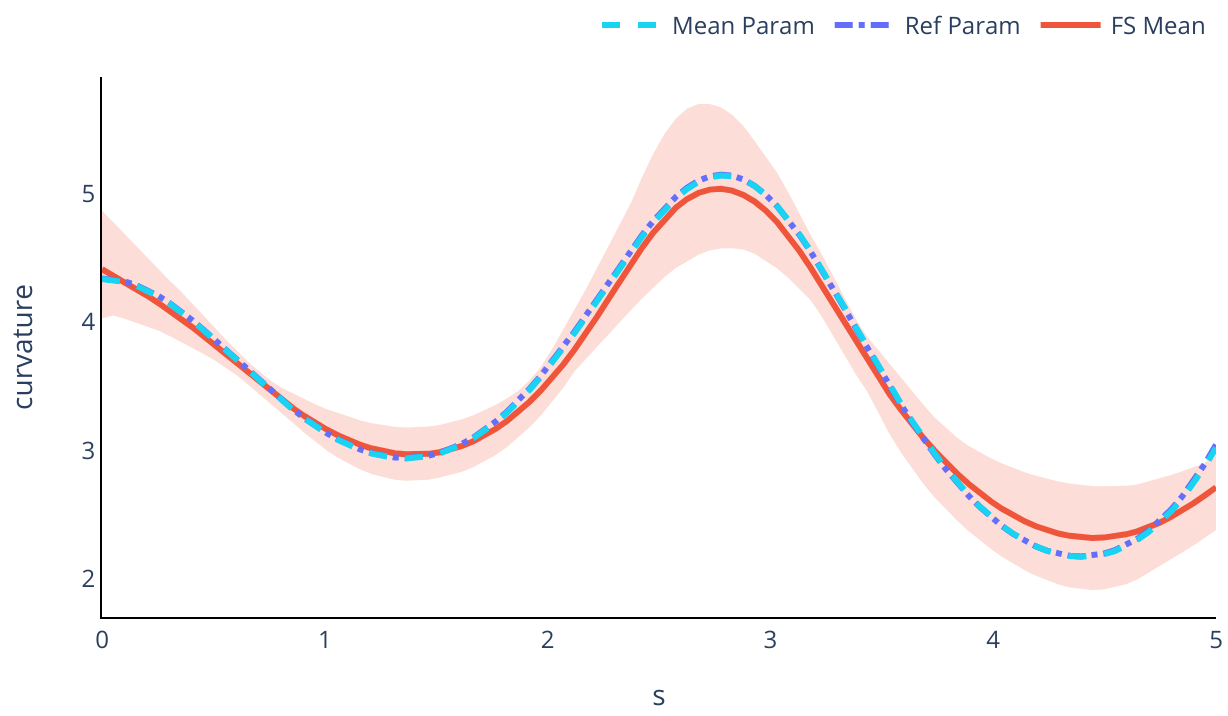} & 
        \includegraphics[width=0.4\textwidth]{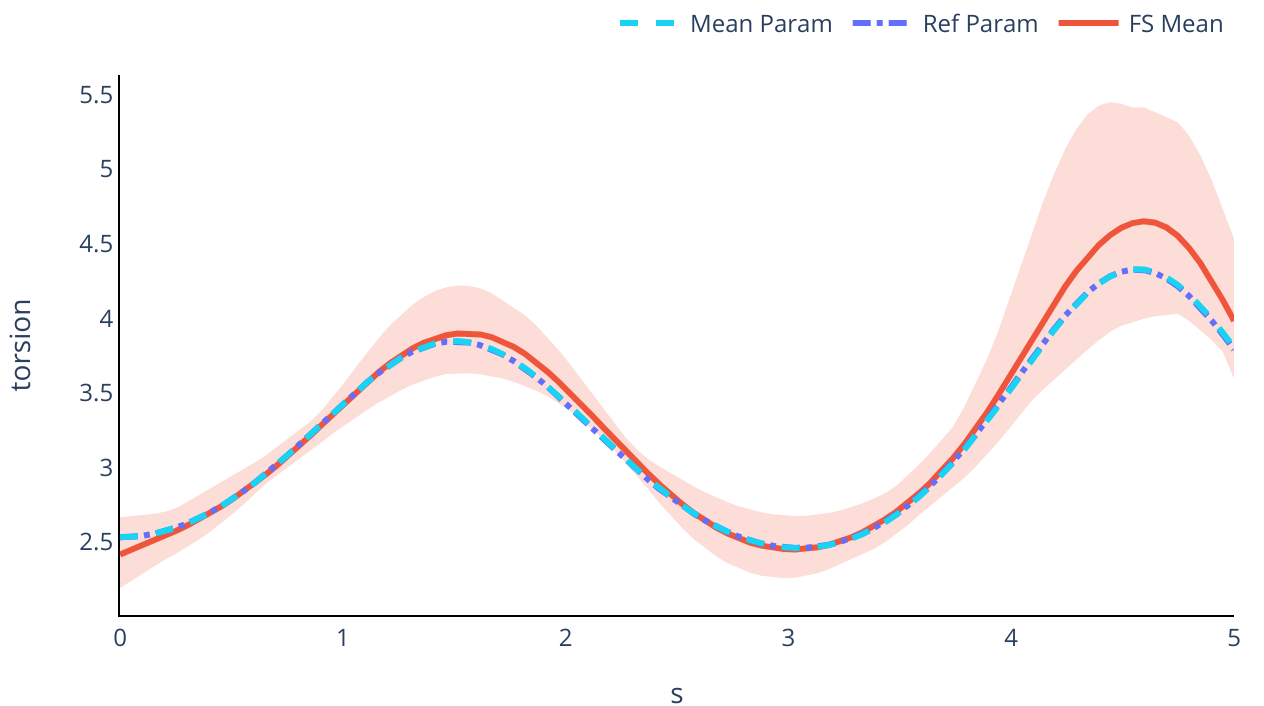} \\
        \multicolumn{2}{c}{\textbf{S3.2}}  \\
        %\label{fig:Fig_S3}
        \multicolumn{2}{c}{
        \includegraphics[width=0.4\textwidth]{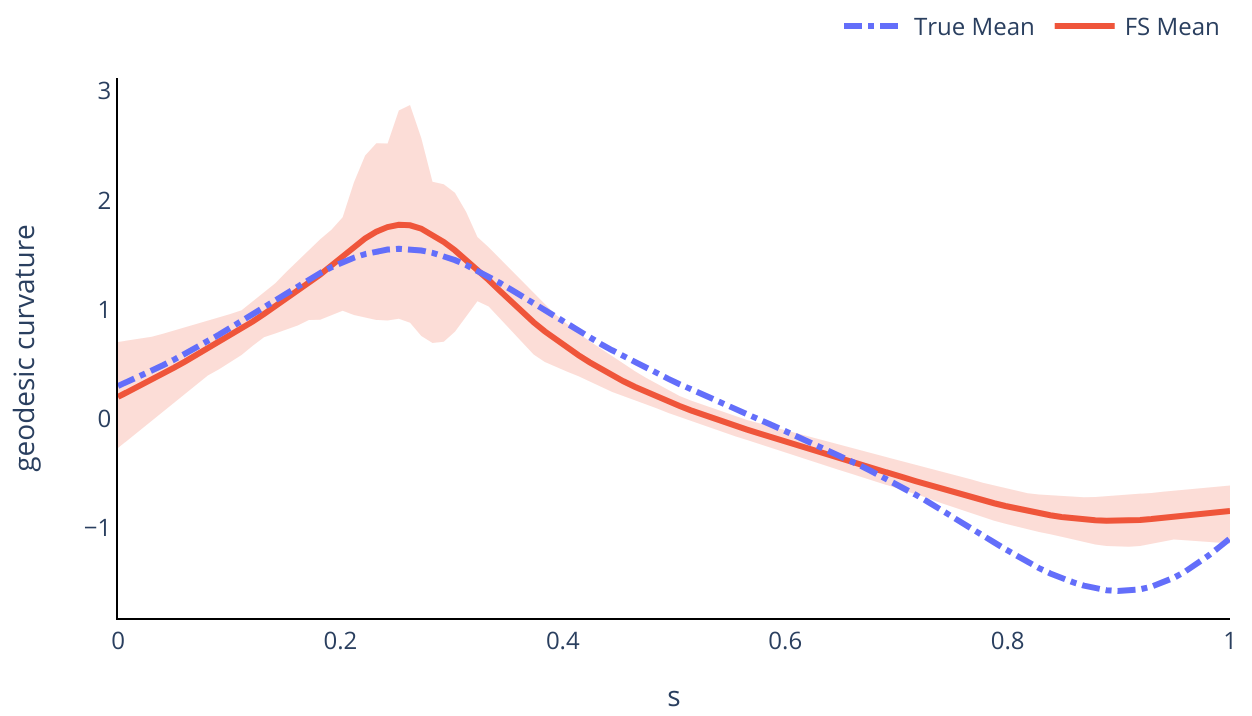}}\\  
        \multicolumn{2}{c}{\textbf{S4}}
        %\label{fig:Fig_S4}
         \end{tabular}
    \caption[]{\label{fig:Fig_Res} Summary of mean parameter estimates under four scenarios without noise over repetitions. True reference mean parameter is in dash-dotted dark blue and the Frenet-Serret (FS) mean in solid red is the average over 100 repetitions. Shaded regions are the maximum and minimum bounds over repetitions. %continuous error bands of Frenet-Serret estimations (shaded red region). 
    For S3 the observed mean parameter is shown in dashed light blue line.}
\end{figure}

The proposed mean parameter estimate is denoted by $\hat\theta^{pop}$.  %and the corresponding Frenet paths by $\hat{Q}^{FS}$ and the mean shape by $\hat X^{FS}$. 
For comparison, we include two alternatives: $\hat{\theta}^{ind}$ %=(1/N)\sum_{i=1}^{N}\hat\theta_i$ 
defined as the average of the individual estimates by the proposed method, and $\hat{\theta}^{ind}_{Ext}$ defined as the median, due to its instability, of the individual estimates computed by extrinsic formulas. 
The results are summarized in Table~\ref{tab:theta} with standard errors in parentheses. Figure~\ref{fig:Fig_Res} shows the average values of the estimates, in comparison to the true mean, over 100 repetitions and the shaded areas represent the maximum and minimum bounds.

%%%%%%%%%%%%%%%%%%%%%%%%%%%%%%%%%%%%%%%%%%%%%%%%%%%%%%%%%%%%%%%%%%%
%%%%%%% Table 1 error on parameters
\begin{table}[tbh]
{\footnotesize
\centering
\hspace*{-0.5cm}\begin{tabular}{|c|c|c|c|c|c|c|c|}
\hline & error & $\| \hat{\kappa}^{ind}_{Ext}-\bar{\kappa}\| _{L^{2}}^{2}$ & $\| \hat{\kappa}^{ind}-\bar{\kappa}\| _{L^{2}}^{2}$ & $\| \hat{\kappa}^{pop}-\bar{\kappa}\| _{L^{2}}^{2}$ & $\| \hat{\tau}^{ind}_{Ext}-\bar{\tau}\| _{L^{2}}^{2}$ & $\| \hat{\tau}^{ind}-\bar{\tau}\| _{L^{2}}^{2}$ & $\| \hat{\tau}^{pop}-\bar{\tau}\| _{L^{2}}^{2}$ 
\tabularnewline\hline\hline 
\multirow{ 2}{*}{\textbf{S1.1}}  & 0  &  \cellcolor{gray!40} & 0.004 (0.003) & 0.004 (0.003) & \cellcolor{gray!40} & 0.004 (0.003) & 0.003 (0.002) \tabularnewline
\cline{2-2}  \cline{4-5} \cline{7-8} 
& 10 & \cellcolor{gray!40} & 0.009 (0.005) & 0.008 (0.004) & \cellcolor{gray!40} & 0.004 (0.003) & 0.004 (0.003) 
\tabularnewline \hline 
\multirow{ 2}{*}{\textbf{S1.2}} & $0$ & 11 (2) & 0.107 (0.076) & 0.102 (0.071) & 0.599 (0.118) & 0.107 (0.081) & 0.089 (0.080) \tabularnewline
\cline{2-8} 
& $0.05$ & 396 (96) & 0.594 (0.188) & 0.605 (0.228) & 18 (7) & 2.744 (2.269) & 2.139 (2.483) \tabularnewline
\hline 
\multirow{ 2}{*}{\textbf{S2.1}} & 0 & \cellcolor{gray!40} & 0.319 (0.065) & 0.021 (0.020) & \cellcolor{gray!40} & 0.495 (0.092) & 0.042 (0.043) \tabularnewline
\cline{2-2}  \cline{4-5} \cline{7-8} 
& 10 & \cellcolor{gray!40} & 1.158 (0.300) & 1.381 (1.456) & \cellcolor{gray!40} & 1.548 (0.382) & 0.573 (0.455) \tabularnewline
\hline 
\multirow{ 2}{*}{\textbf{S2.2}} &  0 & 39 (2) & 0.280 (0.020) & 0.028 (0.020) & 16 (1) & 0.597 (0.005) & 0.099 (0.069) \tabularnewline
\cline{2-8} 
& 0.01 & 61 (12) & 1.260 (0.110) & 1.182 (0.097) & 17 (1) & 1.575 (1.293) & 1.967 (1.500) \tabularnewline
\hline 
\multirow{ 2}{*}{\textbf{S2.3}} & 0 & 40 (2) & 0.346 (0.024) & 0.028 (0.023) & 16 (1) & 0.791 (0.005) & 0.156 (0.063) \tabularnewline
\cline{2-8}
& 0.01 & 326 (79) & 0.608 (0.300) & 0.729 (0.456) & 22 (3) & 1.691 (1.256) & 1.319 (1.355) \tabularnewline
\hline
\multirow{ 2}{*}{\textbf{S3.1}} & 0 & 8.001 (0.177) & 0.018 (0.005) & 0.018 (0.005) & 1.633 (0.033) & 0.015 (0.006) & 0.015 (0.007) \tabularnewline
\cline{2-8} 
& $0.03$ & $5e^{4}$ (193) & 0.520 (0.316) & 0.445 (0.261) & 564 (102) & 3.636 (2.524) & 2.432 (2.211) \tabularnewline
\hline 
\multirow{ 2}{*}{\textbf{S3.2}}& $0$ & 7.785 (0.512) & 0.048 (0.027) & 0.047 (0.027) & 1.686 (0.079) & 0.055 (0.048) & 0.053 (0.048)  \tabularnewline
\cline{2-8} 
& $0.03$ & $5e^{4}$ (253) & 0.583 (0.416) & 0.488 (0.324) & 570 (105) & 3.793 (2.790) & 2.994 (2.790) \tabularnewline
\hline 
\end{tabular}
\caption{\label{tab:theta} Estimation errors on mean parameters.}
}
\end{table}

In most cases the results are better with estimation from Frenet paths than from Euclidean curves, the preprocessing required to estimate Frenet paths from curves adds noise which impacts the results. The difference between the individual and global estimates depends very much on the model of the simulation. For \textbf{S1}, the variability within the population being more additive, there is no big difference between $\hat{\theta}^{ind}$ and $\hat{\theta}^{pop}$, even if the global estimate remains better. On the contrary, in \textbf{S2} with phase warping functions, the results on the case without noise attest to the interest and efficiency of our method with alignment. In noisy cases, it seems that our alignment algorithm suffers, due to the difficulty in alignment with noisy data. This suggests that smoothing methods could be further explored in future development. 
%and is left for future development so in a future development, a first smoothing of the Frenet paths could be applied in order to overcome this.  
The results of \textbf{S3} show the advantage of global estimation over other types of model. Moreover, we observe that for all the scenarios the torsion is a little more difficult to estimate, as it is linked to the third derivative of the curve which is harder to estimate. Finally, even if we use the same non-parametric estimates of the derivatives, the estimates with the extrinsic formula are very unstable whereas they are much more robust with the proposed method as our approach eliminates oscillations and noise more effectively with a joint estimation of $\kappa$ and $\tau$, which makes the overall shape more faithful. 
Additional comparison on the quality of the estimated Frenet paths (Table A in the supplementary) gives a similar conclusion.

\begin{figure}[htbp]
    \centering
    \begin{tabular}{ccc}
        \includegraphics[width=0.28\textwidth]{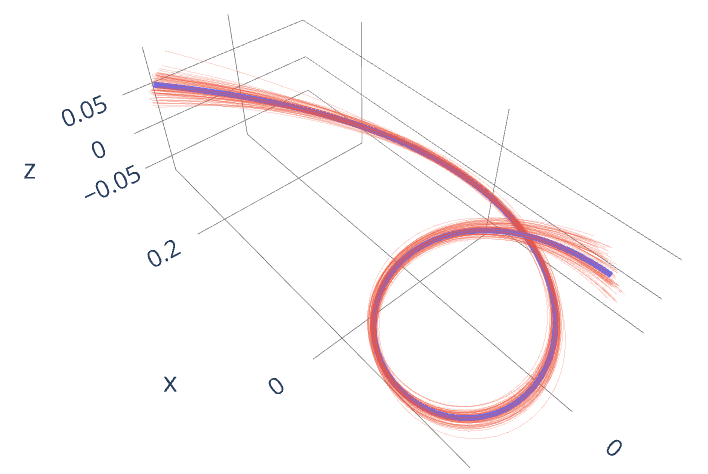} &
        \includegraphics[width=0.28\textwidth]{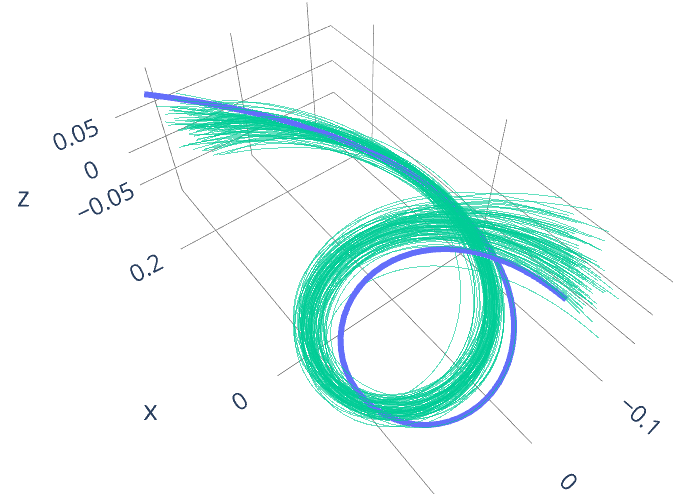} & 
        \includegraphics[width=0.28\textwidth]{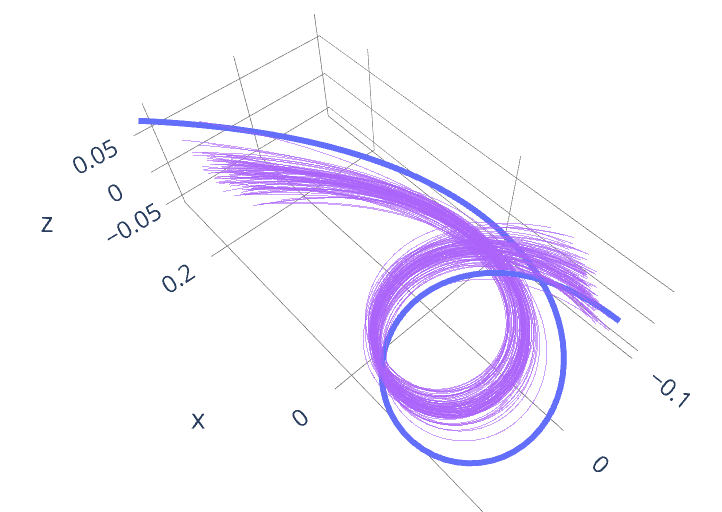} \\
        \multicolumn{2}{c}{\textbf{S1.2}} \\
        \includegraphics[width=0.28\textwidth]{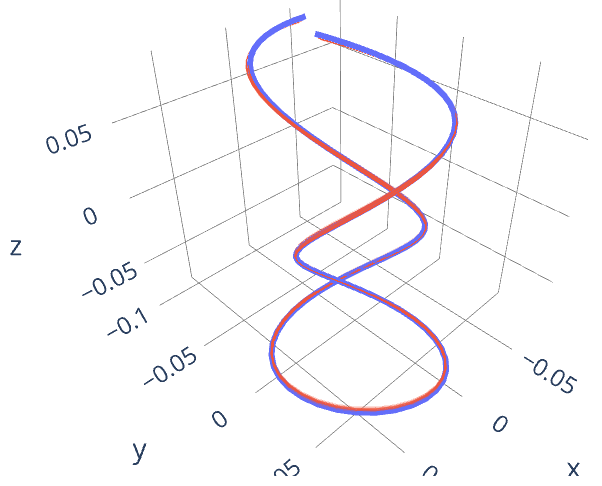} &
        \includegraphics[width=0.28\textwidth]{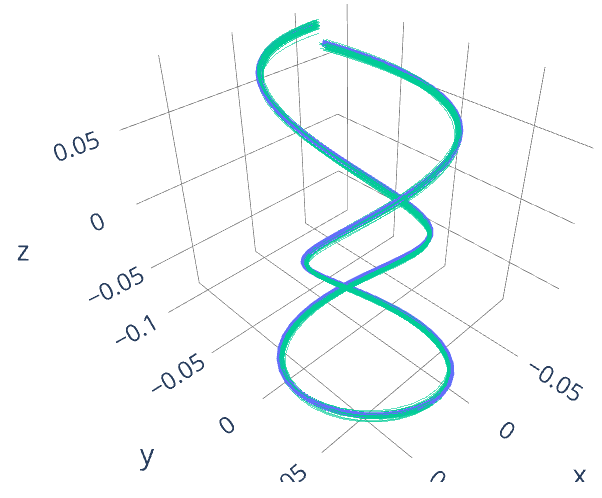} &
        \includegraphics[width=0.28\textwidth]{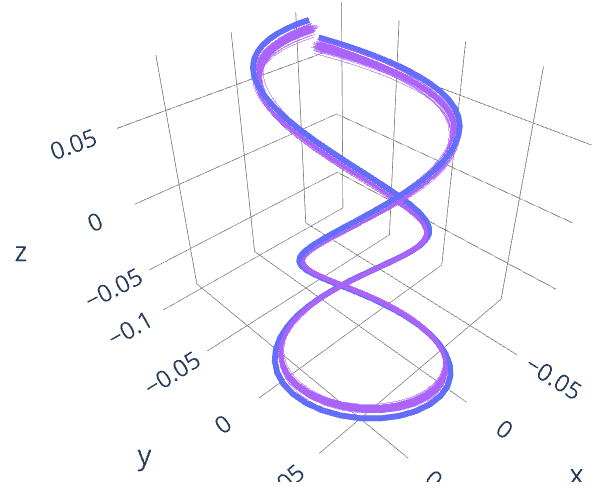} \\
        \multicolumn{2}{c}{\textbf{S2.2}} \\
        \includegraphics[width=0.28\textwidth]{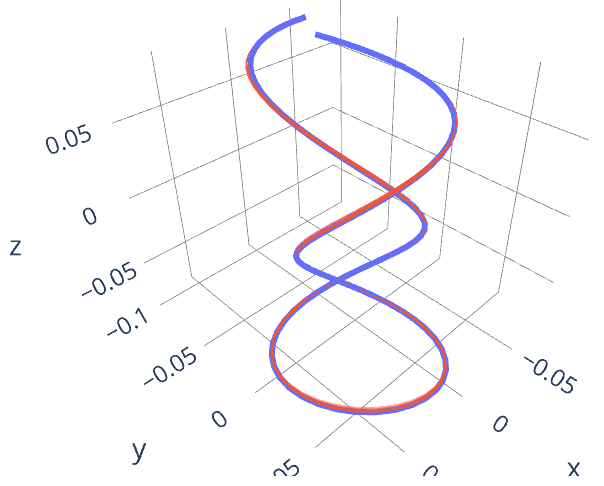} &
        \includegraphics[width=0.28\textwidth]{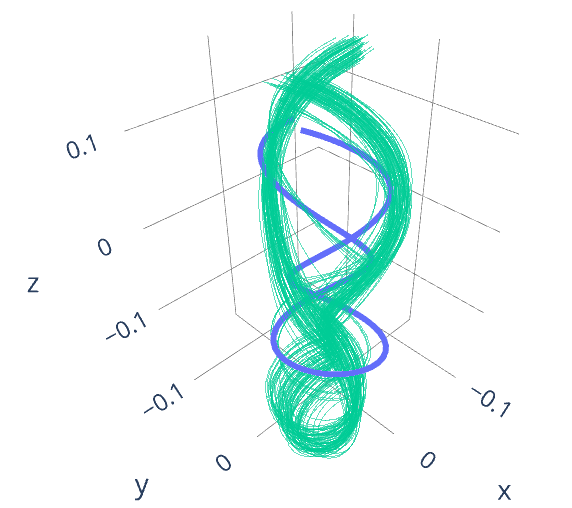} &
        \includegraphics[width=0.28\textwidth]{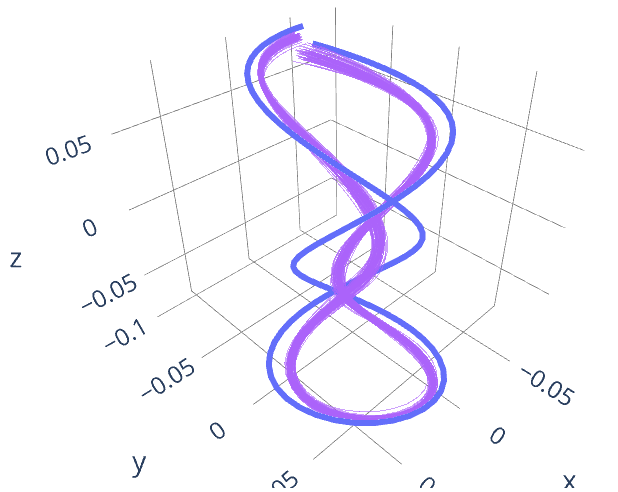} \\
        \multicolumn{2}{c}{\textbf{S2.3}} \\
        \includegraphics[width=0.26\textwidth]{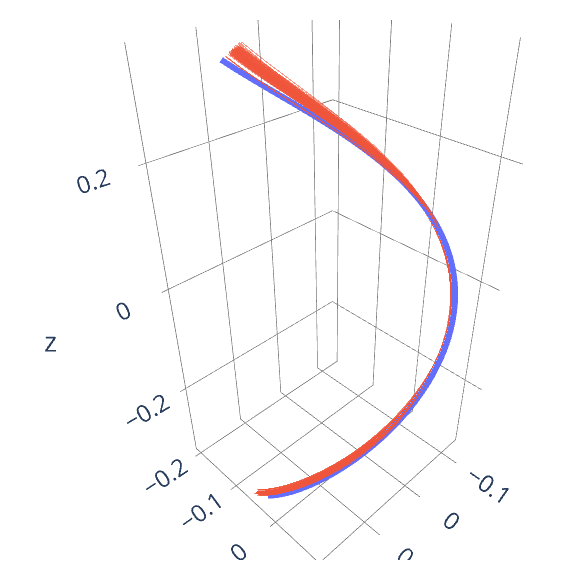} &
        \includegraphics[width=0.26\textwidth]{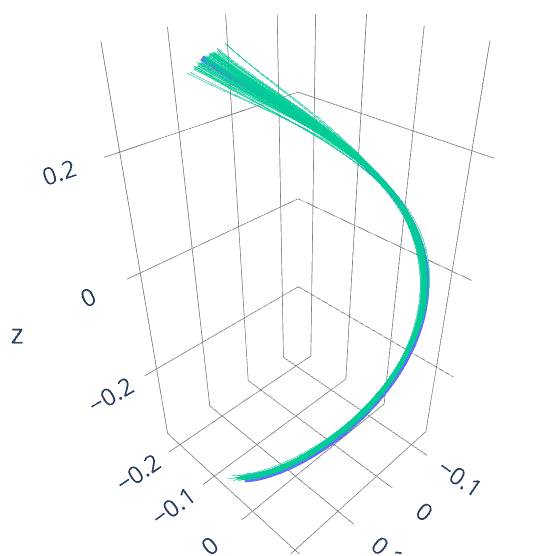} &
        \includegraphics[width=0.26\textwidth]{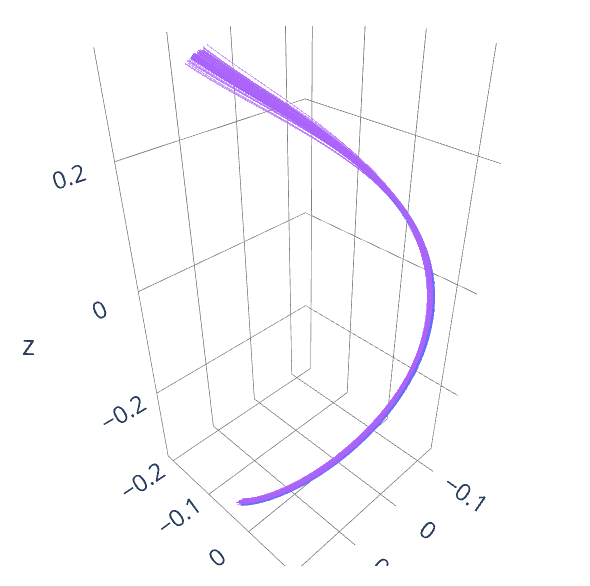} \\
        \multicolumn{2}{c}{\textbf{S3.2}} \\
        \includegraphics[width=0.29\textwidth]{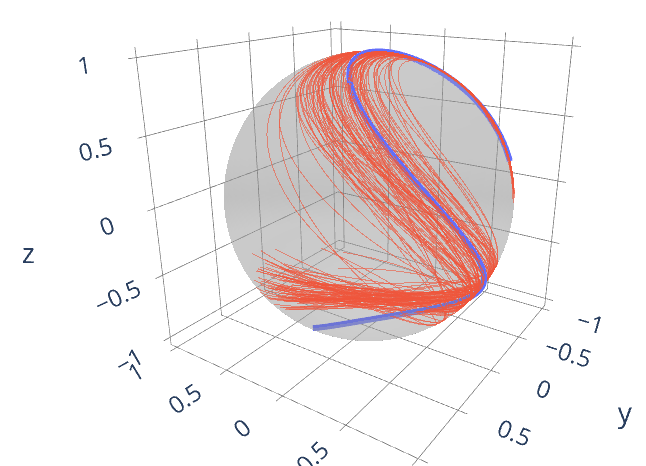} &
        \includegraphics[width=0.29\textwidth]{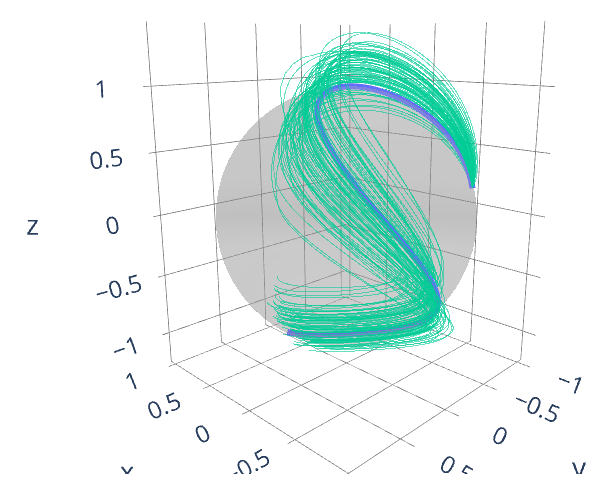} & 
        \includegraphics[width=0.29\textwidth]{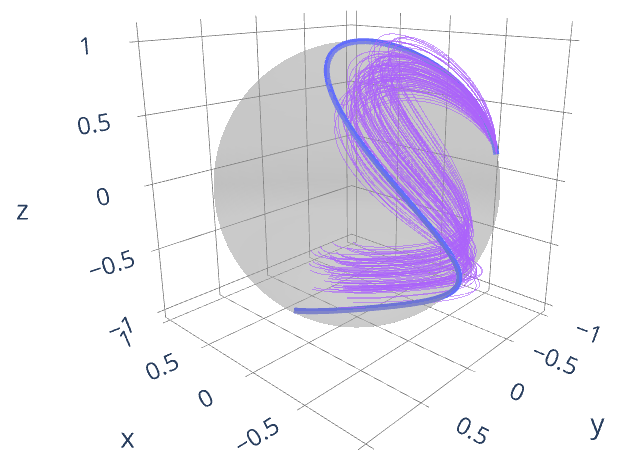} \\
        \multicolumn{2}{c}{\textbf{S4}}
    \end{tabular}
    \caption{\label{fig:Res_EuclideanCurves} Estimation of mean shape under five scenarios over repetitions, with true mean in blue solid line, Frenet-Serret means in the first column (in red), SRVF means in the middle (in green) and Arithmetic means in the last column (in purple).}
\end{figure}

%In addition, we evaluate the quality of the corresponding Frenet paths reconstructed from the estimated mean parameter in terms of the squared geodesic distance to the true Frenet paths relative to that of the initial data.
The proposed mean shape is denoted by $\hat{X}^{FS}$
For comparison, we include the elastic mean by SRVF method $\hat{X}^{SRVF}$ described in section~\ref{sec:elastic} and the arithmetic mean of Euclidean curves $\hat{X}^{Arithm}$. 
%For the quality of the estimation, we compute the $L^2$ distance between the quantities of interest and include Fisher-Rao distance (\ref{eq:DistanceSRVF}) for mean shape. 
The results are visualized in Figure~\ref{fig:Res_EuclideanCurves}.
%For comparison, we include the elastic mean by SRVF method $\hat{X}^{SRVF}$ and the arithmetic mean of the different Euclidean curves $\hat{X}^{Arithm}$. 
Numerical summaries of $L^2$ distance as well as Fisher-Rao distance (\ref{eq:DistanceSRVF}) are compared (Table B. in the supplementary). In addition, for curves on the manifold $\mathbb{S}^2$ (\textbf{S4}) we measure how much the mean belongs to $\mathbb{S}^2$ by $d_{norm}^X = \sum_{j=1}^{n} |\langle X(s_j), X(s_j) \rangle - 1|$ in Table~\ref{tab:theta-S2}.
The $L^2$ distances are very similar between each method for all scenarios. %On the other hand, with the exception of three cases with noisy data (\textbf{S2.2} with noise and \textbf{S3.1, S3.2} with noise), 
The Fisher-Rao distance is comparable, and often smaller with the proposed method than with the SRVF method, even though the latter aims to minimize this distance. The arithmetic method also gives better results with the Fisher-Rao distance than the SRVF method in some cases.
 Overall the distance metrics tend to be similar and do not capture the subtle differences in the geometry very well. 
 Figure~\ref{fig:Res_EuclideanCurves} shows a large difference in results between \textbf{S2.2} and \textbf{S2.3}. Both models are the same except that time warping functions are added in \textbf{S2.3}. Of course, as the points are not distributed in the same way along the curve, this affects the result of the arithmetic mean. In contrast to the SRVF method, our method allows the estimation of time warping and space warping functions separately, and therefore gives much better results in this case. Finally, for spherical curves, the shape seems to be well estimated with the SRVF method but the means are no longer on the sphere, contrary to those estimated by our method, which is clearly seen in Table~\ref{tab:theta-S2}. For an adaptation of SRVF method to manifold data, we refer to \citet{Su2014}.

\begin{table}[htb]
{\small
\centering
\hspace*{-0cm}\begin{tabular}{|c|c|c|c|c|c|c|}
\hline 
   $\sigma_{e}$ & $\| \hat{k}_{g_{Ext}}^{ind}-\bar{k}_g\| _{L^{2}}^{2}$ & $\| \hat{k}_g^{ind}-\bar{k}_g\| _{L^{2}}^{2}$ & $\| \hat{k}_g^{pop}-\bar{k}_g\| _{L^{2}}^{2}$ & $d_{norm}^{FS}$ & $d_{norm}^{SRVF} $ & $d_{norm}^{Arithm}$ \tabularnewline
\hline 
\hline 
$0$ & 0.137 (0.031) & 0.109 (0.043) & 0.093 (0.034) & $1e^{-7}$ ($1e^{-7}$) & 0.172 (0.019) & 0.122 (0.014) \tabularnewline
\cline{1-7} 
 $0.02$ & 0.869 (0.777) & 0.127 (0.043) & 0.109 (0.044) & $3e^{-7}$ ($5e^{-7}$) & 0.181(0.014) & 0.122 (0.013) \tabularnewline
\hline 
\end{tabular}
\caption{\label{tab:theta-S2} Estimation error for spherical curves (S4).}
}
\end{table} 

%%%%%%%%%%%
%% without subfigure %%%%%%%

In terms of computational cost, in the setting of these simulations (25 curves, 100 sample points, 80 iterations of Bayesian optimization and 100 repetitions of each simulation) and with fixed hyperparameters, the estimates ($\hat{\theta}$ and $\bar{X}$) is computed by our algorithm in $1s$ without phase variation and in about $15s$ under phase variation (addition of alignment step), when the SRVF method takes about $7s$ and the Arithmetic method takes $0.1s$. These times are given as an indication insofar as the calculation times of these algorithms depend greatly on the parameters of the simulation. The computation times of our algorithm with fixed parameters seem to be of the same order as those of the compared methods, but our method requires several parameters to be optimised in practice, which considerably increases its computational cost. One iteration of Bayesian optimization with 10 fold cross validation takes about $40s$, so, as we execute all the repetition in parallel, one simulation scenario with all the optimization process takes about one hour. 

%%%%%%%%%%%%%%%%%%%%%%%%%%%%%%%%%%%%%%%%%%%%%%%%%%%%%%%%%%%%%%%%%

\subsection{Real data examples}

We demonstrate our methodology with two different datasets of human movements shown in Figure~\ref{fig:realdata}.
%We compare both the mean parameter ($\theta$) and the mean shape ($\bar X$). For the mean shape, the proposed method with the SRVF mean (\ref{sec:elastic} and the arithmetic mean.
For the observed curves $x_i$, we pre-process the data to create an arclength parametrized data $X_i$ defined on $[0, L_i], i=1,\ldots,n$ and define $Z_i(s) = X_i(sL_i)/L_i, s\in [0,1]$ as a length-normalized curve. 
The raw Frenet paths are obtained from a constrained local polynomial smoothing on the normalized domain $[0,1]$, as $\hat{Q}^{LP}$ in section~\ref{sec:preprocessingX}. 
%The raw Frenet paths are obtained with $\hat{Q}^{LP}$ in section~\ref{sec:preprocessingX}. 

%The mean shape is obtained by solving the ODE with our mean parameters $(\hat{\theta}^{pop})$ and the mean initial condition. As in the simulation we compared this estimate to the one obtained by SRVF method (\texttt{fdasrsf} \cite{fdasrsf}) and to the arithmetic mean of the different Euclidean curves. The latter two means are calculated directly from the time-parameterised curves $x_i(t)$ as the methods do not require a preprocessing step.
%We compare the proposed mean parameter estimates $(\hat{\theta}^{pop})$ with the average of individual estimates by extrinsic formulas (\ref{eq:ExtrinsicCurvatures}) using the local polynomial derivative estimates $(\hat{\theta}_{Ext}^{ind})$. 

% The corresponding mean Euclidean curve (mean shape) is obtained by solving the ODE with our mean parameters $(\hat{\theta}^{pop})$ and the mean initial condition, $\bar{X}^{FS}$. This is compared to the mean curve obtained by SRVF method (\texttt{fdasrsf} \cite{fdasrsf}), $\bar{X}^{SRVF}$, and to the arithmetic mean of the different Euclidean curves, $\bar{X}^{Arithm}$. The latter two means are calculated directly from the time-parameterised curves $x_i(t)$ as the methods do not require a preprocessing step.

%% without subfigure
\begin{figure}[tbhp]
    \centering
    \hspace*{-0.5cm}\begin{tabular}{ccc}
                 \includegraphics[width=0.33\textwidth]{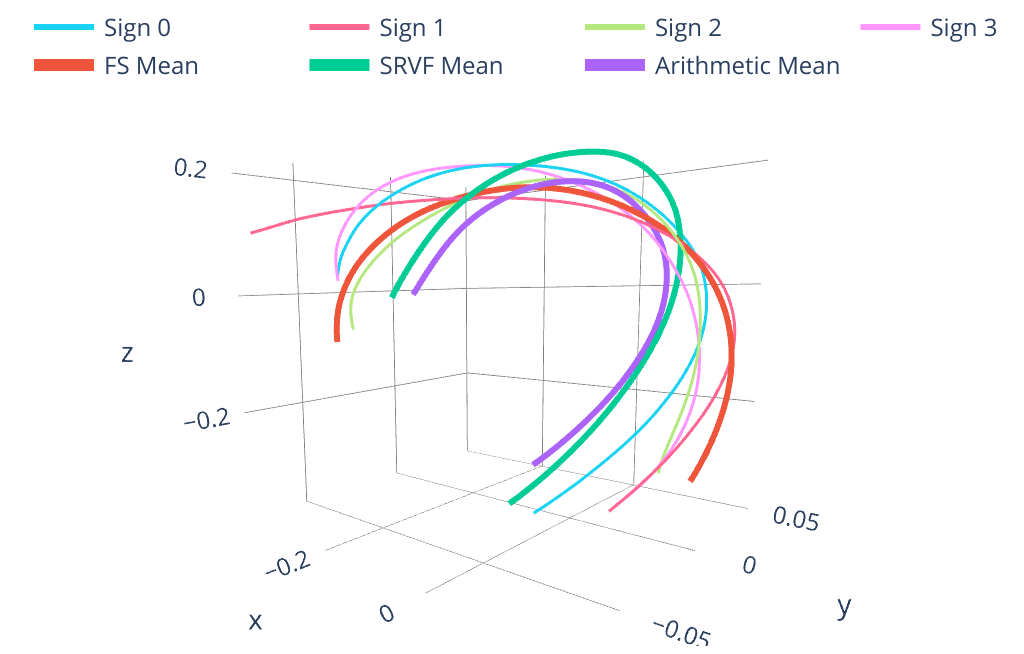} &
         \includegraphics[width=0.33\textwidth]{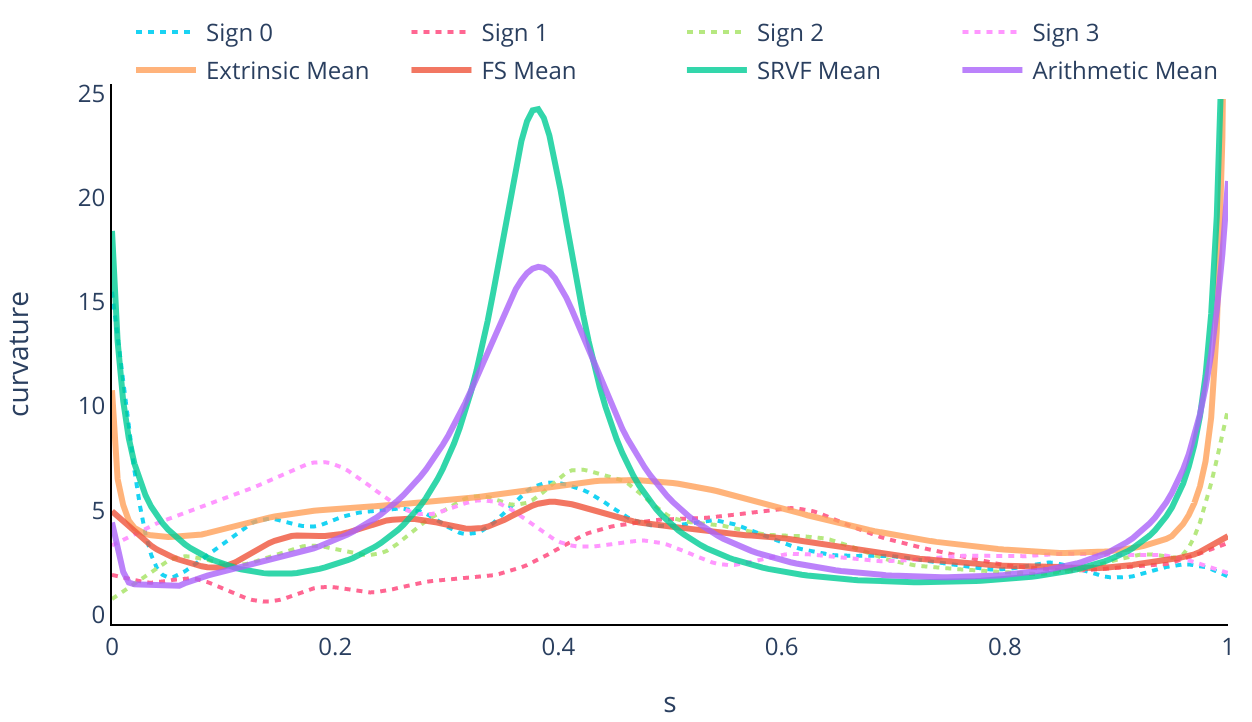} & 
        \includegraphics[width=0.33\textwidth]{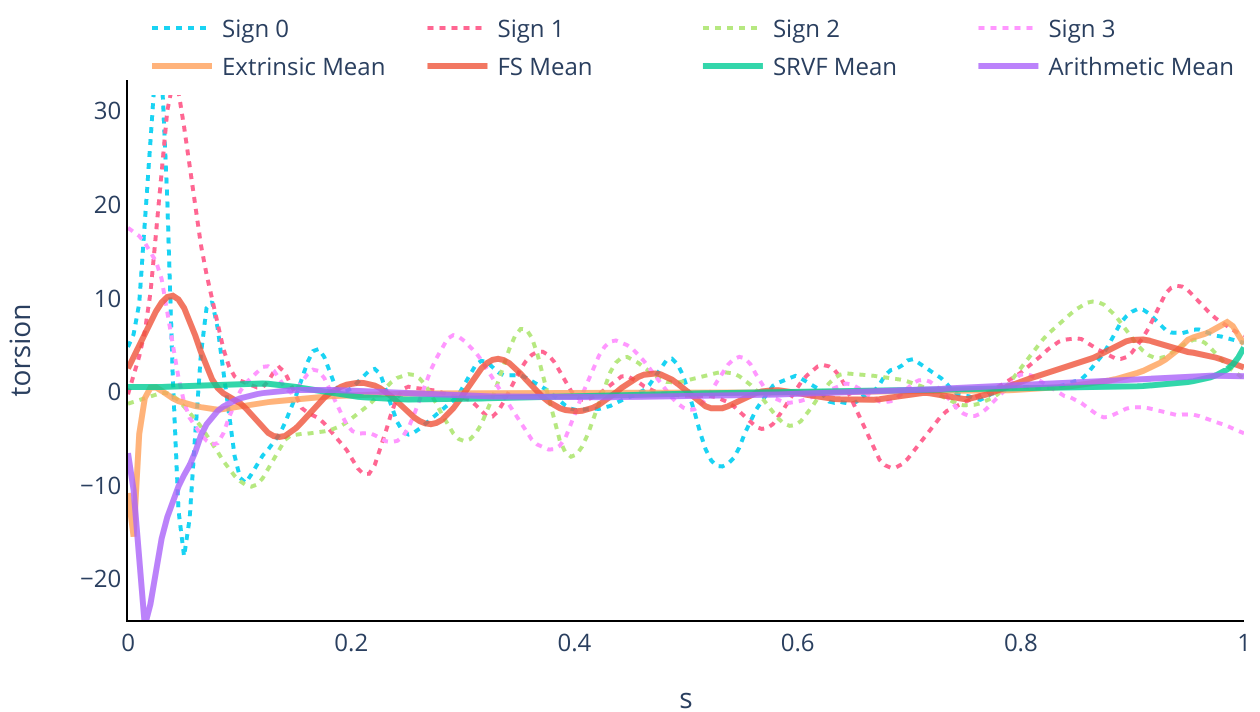} 
    \end{tabular}
    \caption{Analysis of trajectories of sign ``FLY" shown in Figure~\ref{fig:realdata} left. Three mean shape estimates are shown  over scaled trajectories on the left: Frenet-Serret (red), SRVF (green), Arithmetic (purple) means. Frenet-Serret mean curvature (middle) and mean torsion (right) are shown, in comparison with individual estimates. Extrinsic formulas are used to obtain parameters for SRVF and Arithmetic means.}
    \label{fig:FLY}
\end{figure}

%%% using subfigure
%\begin{figure}[tbhp]
%    \centering
%    \begin{subfigure}[b]{1\textwidth}
%        \centering
%        \includegraphics[width=0.35\textwidth]{Figures/FLY/without_smooth_Q/init_curves_0.png}
%        \includegraphics[width=0.45\textwidth]{Figures/FLY/without_smooth_Q/means_0.png}
%        %\subcaption[]{Initial trajectories before scaling (left) and Frenet-Serret (red), SRVF (green) and Arithmetic (purple) mean over the scaled trajectories (right).}
%        %\label{fig:FLY_a}
%    \end{subfigure}
%    \centering
%    \begin{subfigure}[b]{1\textwidth}
%        \centering
%        \includegraphics[width=0.35\textwidth]{Figures/FLY/without_smooth_Q/new_curv_means.png}
%        \includegraphics[width=0.35\textwidth]{Figures/FLY/without_smooth_Q/new_tors_means.png}
%        %\subcaption[]{Curvature (left) and torsion (right) estimates.}
%        %\label{fig:FLY_b}
%    \end{subfigure}
%    \caption{Trajectories of sign ``FLY" in sign language in four repetitions (top left) and three mean shape estimates, Frenet-Serret (red), SRVF (green), Arithmetic (purple) means, over scaled trajectories (top right). Corresponding mean curvature (left) and mean torsion estimates (right) are shown in the bottom, in comparison with individual estimates. Extrinsic formulas are used to obtain parameters for SRVF and Arithmetic means.}
%    \label{fig:FLY}
%\end{figure}
%
%%
\subsubsection{Mocaplab data: Sign "Fly"}
\label{sec:FLY}

The data set shown in the left of Figure~\ref{fig:realdata} consists of four repetitions of the sign "Fly" in American Sign Language by the same deaf signer, collected by the company MOCAPLAB. 
%The acquisition of these data is done by motion capture by the company MOCAPLAB\footnote{https://www.mocaplab.com/fr/}. 
In recent years, the company has developed a very precise technique for acquiring finger and hand movements. These data are therefore low-noise. Movements of 3 points on the right hand are recorded and the trajectory of the barycenter of these points constitutes our original data before scaling, visible on the top left of Figure~\ref{fig:FLY}.
%The raw Frenet paths are obtained by constrained local polynomial smoothing ($Q^{LP}$) with a common bandwidth ($h_1=0.1$).
For estimation, the common bandwidth chosen for estimating the raw Frenet paths from constrained local polynomial smoothing is $h_1= 0.1$ and the hyperparameters are selected from $h\in\left(0.016,0.1\right)$ and $\lambda_{1},\lambda_{2}\in\left(1e^{-10},1e^{-6}\right)$. 

Figure~\ref{fig:FLY} shows three mean shape estimates over the scaled initial data in the left with the parameter estimates in the middle and right. Each Euclidean curve is centred according to its geometric center and the optimal rotation with respect to a reference curve chosen, calculated by Procrustes analysis. On this plot, the SRVF and Arithmetic means appear to have a rather different shape from the replicates and from our mean shape estimate (Frenet-Serret mean). The mean Fisher-Rao distance between the estimated mean curve and each initial scaled curve is $0.091$ ($0.008$) for Frenet-Serret mean, $0.285$ ($0.058$) for SRVF mean and $0.271$ ($0.055$) for the Arithmetic mean (resp. for $L^2$ distance, $0.031$ $(0.007)$ for Frenet-Serret, $0.044$ $(0.009)$ for SRVF and $0.044$ $(0.007)$ for Arithmetic).
The corresponding mean curvature and mean torsion $(\hat{\theta}^{pop})$ are plotted, over the individual estimates $\hat{\theta}_i$ and the mean of the individual extrinsic estimates $(\hat{\theta}_{Ext}^{ind})$. For comparison, we also add the curvatures and torsions of the SRVF and the Arithmetic means computed by extrinsic formulas.
The curvature profiles confirm the observation already made on the Euclidean curves. Indeed, the curvatures of the SRVF and the arithmetic mean show a broad peak that is not present in the individual curvatures and the torsion curves are quite flat compared to the individual ones, while the curvature and torsion of the Frenet-Serret mean show variations much more similar to the individual curves. This is similar to the example case with torsion variability in Section~\ref{sec:comparison}. 

% without subfigure
\begin{figure}[!h]
        \centering
    \hspace*{-0.5cm}\begin{tabular}{ccc}
         \includegraphics[width=0.36\textwidth]{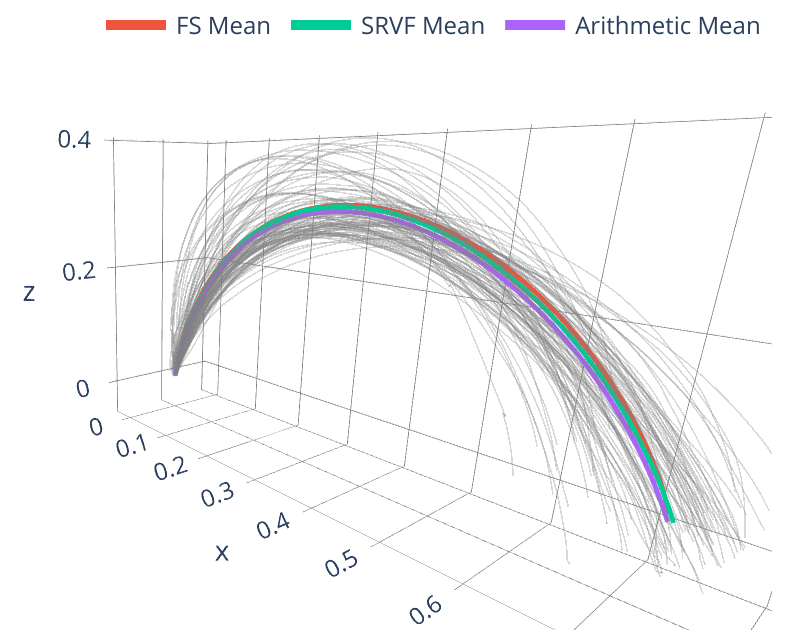} &
        \includegraphics[width=0.35\textwidth]{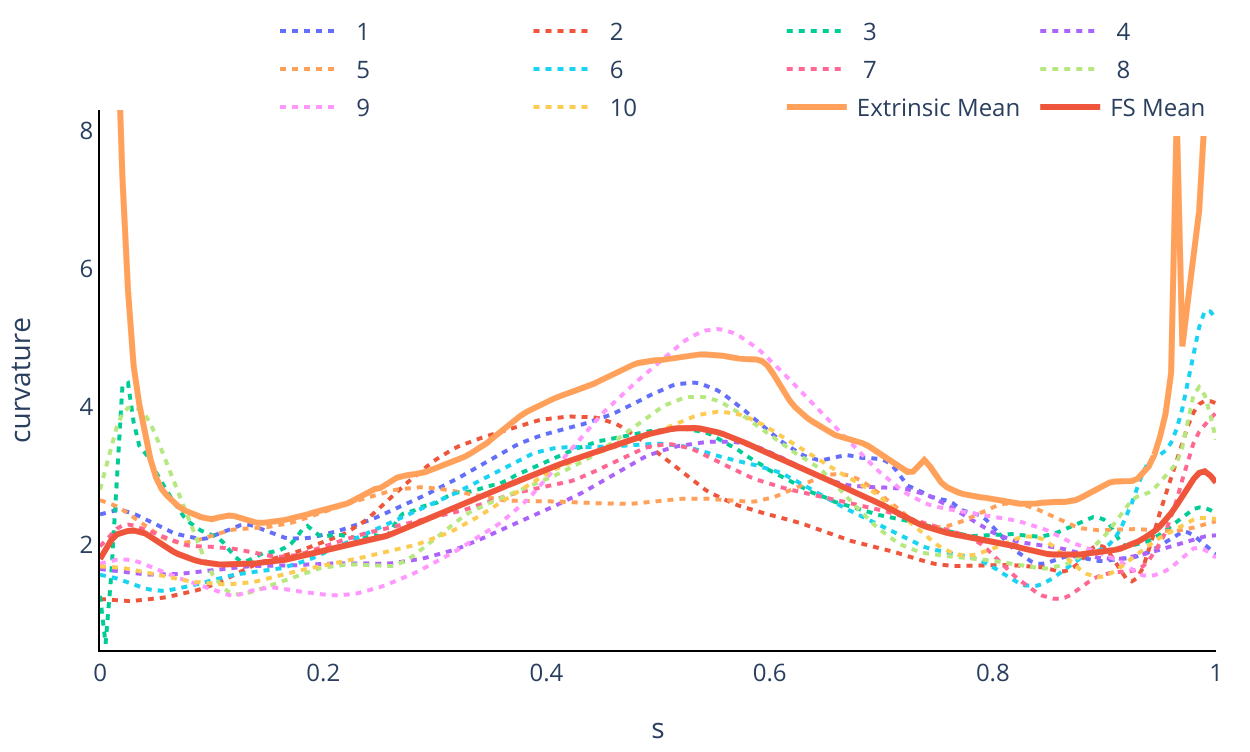} &
        \includegraphics[width=0.35\textwidth]{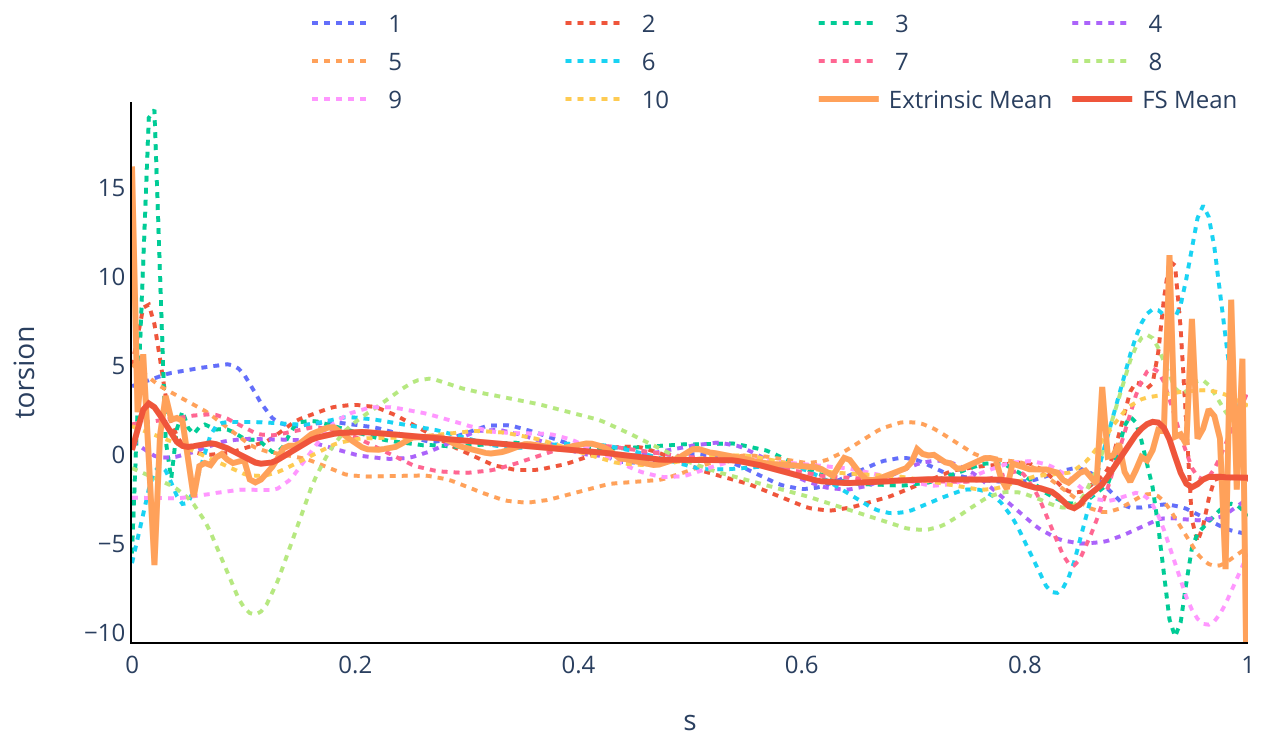} 
    \end{tabular}
    \caption{\label{fig:Raket_specialCase} Analysis of trajectories of hand movement shown in Figure~\ref{fig:realdata} right. Results are for 10 participants with 10 repetitions under Condition 8. Three mean shape estimates, Frenet-Serret (red), SRVF (green) and Arithmetic (purple) means are shown over scaled curves (left). Frenet-Serret mean curvature (middle) and mean torsion (right) per participant are shown, in comparison to individual estimates (dotted) and the extrinsic mean estimates (orange).}
\end{figure}

%
%%% using subfigure
%\begin{figure}[!h]
%    \centering
%    \begin{subfigure}[b]{1\textwidth}
%        \centering
%        \includegraphics[width=0.45\textwidth]{Figures/Raket/without_smooth_Q/init_curves_0.png}
%        \includegraphics[width=0.36\textwidth]{Figures/Raket/without_smooth_Q/means_0.png}
%        %\subcaption[]{10 repetitions of the 10 participants. Left: initial curves (one colour per participant), right: scaled curves and three estimated means over all.}
%        %\label{fig:Raket_specialCase_a}
%    \end{subfigure}
%    \centering
%    \begin{subfigure}[b]{1\textwidth}
%        \centering
%        \includegraphics[width=0.35\textwidth]{Figures/Raket/without_smooth_Q/new_curv_case8.png} \hspace{1cm}
%        \includegraphics[width=0.35\textwidth]{Figures/Raket/without_smooth_Q/new_tors_case8.png}
%        %\subcaption[]{Curvature (left) and torsion (right) means per participant (dotted), mean over all (solid red) and mean of extrinsic estimates (solid orange).}
%        %\label{fig:Raket_specialCase_b}
%    \end{subfigure}
%    \caption{Trajectories of hand movement of 10 participants with 10 repetitions under Condition 8 (top left) and three mean shape estimates, Frenet-Serret (red), SRVF (green) and Arithmetic (purple) means over scaled curves (top right). Frenet-Serret mean curvature (left) and mean torsion (right) per participant are shown in the bottom, in comparison to individual estimates (dotted) and the extrinsic mean estimates (orange).}
%    \label{fig:Raket_specialCase}
%\end{figure}

\subsubsection{Raket et al. data}

The data set shown in the right of Figure~\ref{fig:realdata} are from a biomedical experiment on hand movement in \citet{RaketMarkussen2016}. An experiment is designed to require each participant to move a hand-held object to a target location while avoiding an obstacle. The trajectories of the (three-dimensional) arm movement of each participant are recorded under various experimental conditions, with an aim to characterize the commonality and variations. 

%Fifteen obstacle avoidance tasks were performed (one for every combination of obstacle height, Small (20 cm), Medium (27.5 cm) and Tall (35 cm) and obstacle distance from a starting position (15 cm, 22.5 cm, 30 cm, 37.5 cm, 45 cm) as well as a control experiment with no obstacle. The ten participants repeated each task ten times.
For each condition, we estimate the mean over the 10 different participants and their 10 repetitions. A common bandwidth ($h_1 = 0.2$) is chosen to obtain the raw Frenet paths from constrained local polynomial smoothing. The hyperparameters are selected from $h\in\left(0.01,0.1\right)$ and $\lambda_{1},\lambda_{2}\in\left(1e^{-9},1e^{-7}\right)$. 

%% without subfigure
\begin{figure}[!h]
    \centering
   \hspace*{-0.5cm}\begin{tabular}{ccc}
        \includegraphics[width=0.3\textwidth]{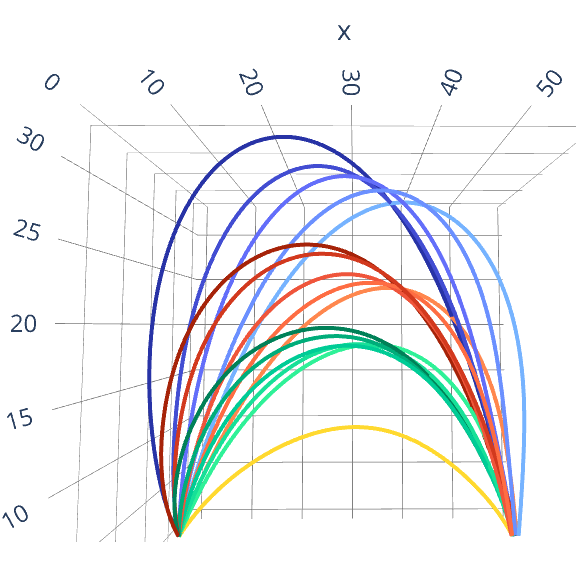} &
        \includegraphics[width=0.34\textwidth]{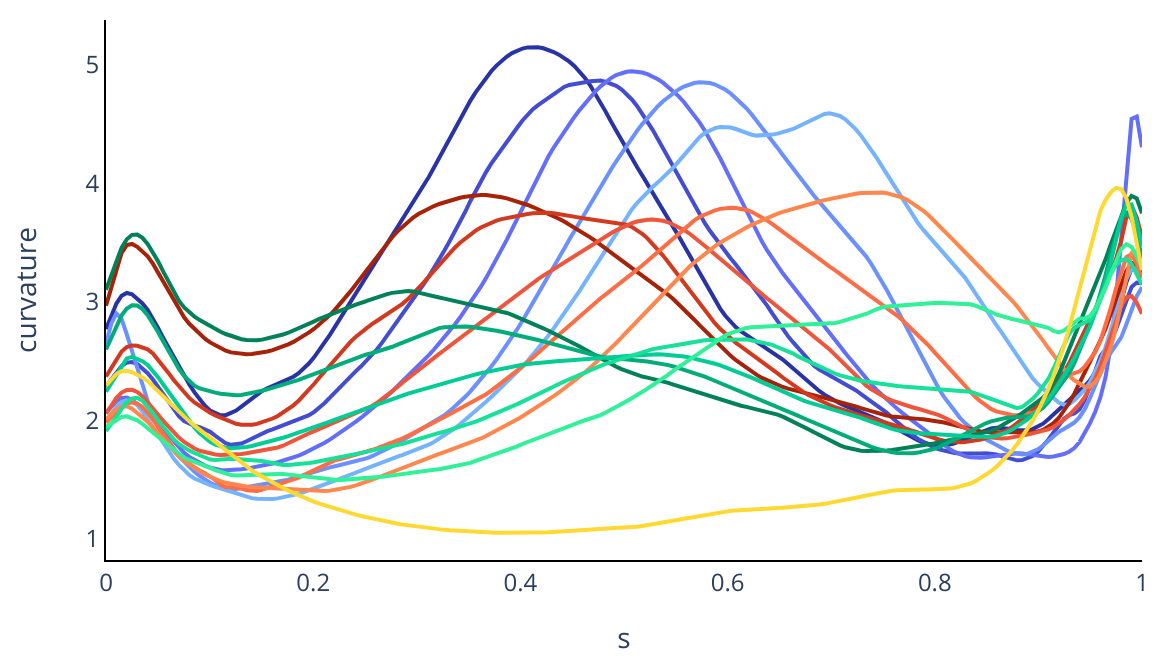} &
        \includegraphics[width=0.34\textwidth]{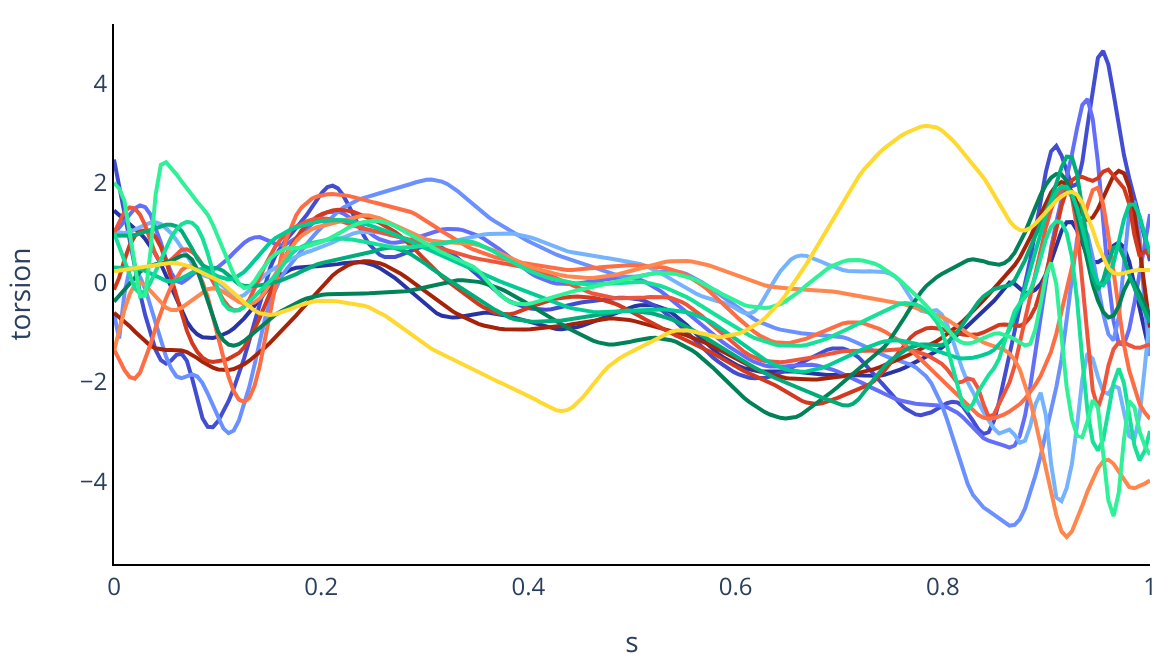} \\
        \multicolumn{3}{c}{\includegraphics[width=0.7\textwidth]{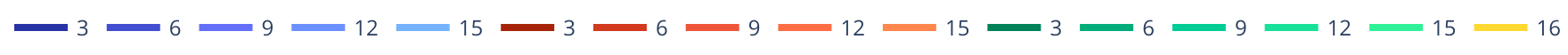}} 
    \end{tabular}
    \caption{\label{fig:Raket_MeansCond} Comparison of mean estimates per condition for hand movement trajectories: mean shape (left), mean curvature (middle) and mean torsion (right). Conditions with tall obstacle are plotted in blue, with medium obstacle in red, small one in green and control condition (without obstacle) in yellow.}
\end{figure}

%%% using subfigure
%\begin{figure}[!h]
%    \centering
%    \begin{subfigure}[b]{1\textwidth}
%        \centering
%        \includegraphics[width=0.3\textwidth]{Figures/Raket/without_smooth_Q/means_cond_000.png}
%        \includegraphics[width=0.34\textwidth]{Figures/Raket/without_smooth_Q/curv_cond_0.png}
%        \includegraphics[width=0.34\textwidth]{Figures/Raket/without_smooth_Q/tors_cond_0.png}
%    \end{subfigure}
%    \centering
%    \begin{subfigure}[b]{1\textwidth}
%        \centering
%        \includegraphics[width=0.7\textwidth]{Figures/Raket/without_smooth_Q/legend_cond.png}
%    \end{subfigure}
%    \caption{Comparison of mean estimates per condition for hand movement trajectories: mean shape (left), mean curvature (middle) and mean torsion (right). Conditions with tall obstacle are plotted in blue, with medium obstacle in red, small one in green and control condition (without obstacle) in yellow.}
%    \label{fig:Raket_MeansCond}
%\end{figure}

Figure~\ref{fig:Raket_specialCase} presents results from one representative case (Medium obstacle at distance 30.0 cm), of 10 repetitions of 10 participants (one colour per participant). The same data, scaled, are visible in grey on the left of Figure~\ref{fig:Raket_specialCase}, on which the three estimated mean shapes (SRVF, Arithmetic, Frenet-Serret) have been displayed. On the middle and right are shown the proposed mean curvature and mean torsion estimates (red solid), over the mean per participant (dotted). We add the mean of extrinsic estimates (orange solid) for comparison. It can be seen that there is not much variation in amplitude and phase between the different mean parameters per participant. This could explain why the three estimated means are very similar in shape. 

% The third column shows the estimated mean curves (SRVF, Arithmetic, Frenet-Serret), plotted over all samples.

In Figure~\ref{fig:Raket_MeansCond} we compare the proposed mean estimates (Euclidean curve, curvature, torsion) for each condition, over all subjects and their repetitions. 
It appears that the curvatures reflect very well the different distances and heights of the obstacles in each condition, while the torsions are rather similar across all conditions and do not allow to differentiate the three heights for example. We observe a huge difference on curvature and torsion plot, between the control condition done without any obstacle in yellow and the others. This example shows the interest of a method for estimating the mean geometry as well as the mean shape. These additional estimates and information could be used in a complex model of variance analysis \citep[e.g.,][]{Backenroth2018} and this would be an interesting direction to explore for future work.

\bibliography{Paper1_SerretFrenet}

\end{document}